\documentclass{article}
\usepackage{emulateapj_perso}
\usepackage{apjfonts}
\usepackage{psfig}

\def\ltsima{$\; \buildrel < \over \sim \;$} 
\def\gtsima{$\; \buildrel > \over \sim \;$}
\def\proptosima{$\; \buildrel \propto \over \sim \;$}
\def\simlt{\lower.5ex\hbox{\ltsima}}
\def\simgt{\lower.5ex\hbox{\gtsima}}
\def\simpropto{\lower.5ex\hbox{\proptosima}}

\makeatletter
\let\@internalcite\cite
\def\cite{\def\astroncite##1##2{##1\ ##2}\@internalcite}
\def\citey{\def\astroncite##1##2{##1\ (##2)}\@internalcite}
\def\@citex[#1]#2{\if@filesw\immediate\write\@auxout{\string\citation{#2}}\fi
  \def\@citea{}\@cite{\@for\@citeb:=#2\do
    {\@citea\def\@citea{; }\@ifundefined
       {b@\@citeb}{{\bf ??}\@warning
       {Citation `\@citeb' on page \thepage \space undefined}}
{\csname b@\@citeb\endcsname}}}{#1}}
\def\@cite#1#2{#1\if@tempswa #2\fi}
\def\@biblabel#1{}
\def\astroncite#1#2{#1\ #2}
\makeatother

\begin{document}
\lefthead{Montmerle et al.}
\righthead{Rotation and X-ray emission from protostars}

\title{Rotation and X-ray emission from protostars}

\author{
Thierry Montmerle\altaffilmark{1} AND Nicolas Grosso\\
{\scriptsize \rm Service d'Astrophysique, CEA/DAPNIA/SAp, Centre d'Etudes de Saclay, 
91191 Gif-sur-Yvette Cedex, France}\\
Yohko Tsuboi\\
{\scriptsize \rm Department of Astronomy \& Astrophysics, Penn State University,
University Park, PA 16802, USA}\\
AND\\
Katsuji Koyama\altaffilmark{2}\\
{\scriptsize \rm Department of Physics, Faculty of Science, Kyoto University,
Sakyo-ku, Kyoto 606-8502, Japan}\\
{\it \scriptsize Received 1999 July 12; accepted 1999 November 18}}
\altaffiltext{1}{\it Send offprint request to {\tt montmerle@cea.fr}.}
\altaffiltext{2}{CREST, Japan Science and Technology Corporation
(JST), 4-1-8 Honmachi, Kawaguchi, Saitama, 332-0012, Japan}

\begin{abstract} The {\sl ASCA} satellite has recently detected variable hard
X-ray emission from two Class~I protostars in the $\rho$ Oph cloud, YLW15
(IRS43) and WL6, with a characteristic time scale $\sim 20$h.  In YLW15, the
X-ray emission is in the form of quasi-periodic energetic flares, which we
explain in terms of strong magnetic shearing and reconnection between the
central star and the accretion disk.  The flare modelling, based on the solar
analogy, gives us access to the size of the magnetic structures, which in turn
allows to calculate the rotation parameters of the star and the disk.  In WL6,
X-ray flaring is rotationally modulated, and appears to be more like the
solar-type magnetic activity ubiquitous on T Tauri stars.  On the basis of these
observations, we find that YLW15 is a fast rotator (near break-up), while WL6
rotates with a significantly longer period.  We thus use X-ray flaring as a
``clock'' to measure the rotation of protostars.  With the help of the
mass-radius relation on the stellar ``birthline'', we derive a mass $M_\star
\sim 2 M_\odot$ and $\simlt 0.4 M_\odot$ for the central stars of YLW15 and WL6
respectively.  YLW15 thus appears as a future A star.  On the long term, the
magnetic interactions between the star and the disk results in magnetic braking
and angular momentum loss of the star.  The compared rotation behavior of YLW15
and WL6 confirms that for solar-mass stars their magnetic braking takes place on
time scales $t_{br} \sim$ a few $10^5$\,yrs, i.e., of the same order as the
estimated duration of the Class~I protostar stage.  The main parameter
determining $t_{br}$ turns out to be the stellar mass, so that close to the
birthline there must be a mass-rotation relation, $t_{br} \simpropto M_\star$,
such that stars with $M_\star \simgt 1 - 2$\,M$_\odot$ are fast rotators, while
their lower-mass counterparts have had the time to spin down and reach
synchronous rotation with the inner surrounding accretion disk.  The rapid
rotation and strong star-disk magnetic interactions of YLW15 also naturally
explain the observation of ``superflares'' of X-ray luminosities as high as $L_X
\simgt 10^{33-34}$\,erg\,s$^{-1}$ during a few hours, while at the WL6 stage the
lower X-ray luminosities are likely to be of purely stellar origin.  The
mass-rotation relation through magnetic braking may also explain why so few
Class~I protostars have been detected in X-rays to date, and why they all lie in
clusters.  In the case of YLW15, and perhaps also of other protostars, a hot
coronal wind ($T \sim 10^6$ K) may be responsible for the VLA thermal radio
emission.  This paper thus proposes the first clues to the magnetic properties
of protostars, which govern their rotation status and evolution.


\end{abstract}
\vspace{0.75cm}

\section{Introduction}

Using the {\sl ASCA} satellite, Tsuboi et al.  (1999; hereafter Paper I) have
recently detected approximately periodic (characteristic time scale $\sim
20$\,h) X-ray emission from one remarkable protostar in the nearby ($d \sim
160$\,pc) \footnote{The 165\,pc value from Dame et al.  1987 was adopted in
paper~I to estimate the intrinsic X-ray luminosities.}  $\rho$ Ophiuchi cloud
core region, YLW15 (Young, Lada, \& Wilking 1986), also known as IRS43 (Wilking,
Lada, \& Young 1989), at a level of $L_X = 20-5 \times 10^{31}$\,erg\,s$^{-1}$.
This protostar belongs to the so-called ``Class~I'' stage, characterized by a
strong mid-IR excess, and corresponds to the final phases of the accretion
process, where most of the mass is contained in the central star, with
comparatively little left in the circumstellar envelope.  The infalling material
transits mostly via an accretion disk (e.g., Andr\'e \& Montmerle 1994:  AM;
Andr\'e, Ward-Thompson \& Barsony 2000:  AWB).  YLW15 displayed quasi-periodic
X-ray flares; prior to this observation, YLW15 was seen by the {\sl ROSAT} High
Resolution Imager (HRI) to emit a ``superflare'' with a peak X-ray luminosity
$L_X \simgt 10^{33-34}$\,erg\,s$^{-1}$ during a few hours (Grosso et al.  1997).

The main properties of YLW15 can be summarized as follows.  Its bolometric
luminosity is $L_{bol} \sim 10$\,L$_\odot$ (Wilking, Lada \& Young 1989), making
it one of the most luminous protostars in the $\rho$ Oph Core F region.  It is
surrounded by a relatively dense, compact dusty envelope with outer radius
$R_{out}\simlt$3000\,AU, and total (gas~$+$~dust) mass $M_{env}$= 0.05--0.3\,
$M_{\odot}$ (AM; Motte, Andr\'e, \& Neri 1998:  MAN).  The slope of its infrared
(IR) spectral energy distribution (SED), $\alpha_{IR} = d({\rm log}\lambda
F_{\lambda})/d{\rm log}\lambda = 1.2$, led to its early classification as a
Class~I protostar.  From $JHKL$ photometry, one can compute the extinction,
which suffers a large uncertainty:  $A_{V,IR} \sim 20 - 40$, with a
``preferred'' value $A_{V,IR} \sim 30$, or equivalently $N_H \sim 6.7 \times
10^{22}$\,cm$^{-3}$ (see discussion in Grosso et al.  1997).  The X-ray spectral
fitting yields $A_{V,X} \sim 15$ (Paper I), which can be considered as an
acceptable agreement considering the rather large uncertainties on $A_{V,IR}$;
alternatively, this difference could be due to a smaller gas-to-dust ratio in
dense molecular cores.  YLW15 also powers a compact bipolar CO outflow (Bontemps
et al.  1996), and exhibits a circumstellar disk $\simgt 500$\,AU in radius,
seen approximately edge-on in $HST-NICMOS$ images (Terebey et al.  1999, in
preparation).  In addition, with a VLA 6-cm flux density of 3.3\,mJy, it is by
far the brightest stellar radio continuum source of Core F (Leous et al.  1991),
which is likely of thermal origin (Andr\'e et al.  1992).  YLW15 is thus
confirmed to be a bona-fide Class~I protostar, at an intermediate stage of
evolution between the younger ``Class 0'' protostars (Andr\'e, Ward-Thompson, \&
Barsony 1993, AWB) and ``flat-spectrum'' protostars (AM).  The estimated age of
such protostars is $\sim 0.75-1.5 \times 10^5$\,yr (Luhman \& Rieke 1999).

In this paper, we argue that the quasi-periodic X-ray flares of YLW15 can be
explained by fast rotation of the central star with respect to the inner
accretion disk, which results in star-disk shearing of the magnetic field lines,
producing magnetic reconnection and mass loss, and eventually extremely high
X-ray luminosities.  On the other hand, we show that magnetic braking, due to
the coupling between the star and the disk, occurs on timescales typically
$\simlt$ a few $10^5$\,yrs, i.e., on the order of the age of Class~I protostars.
This braking asymptotically leads to synchronous rotation between the central
star and the inner accretion disk (period $\sim$ a few days).  Comparing
the X-ray light curves of YLW15 and of WL6, another Class~I protostar in the
$\rho$ Oph core F showing rotational modulation in X-rays corresponding to
a period $\sim 3$\,days, we argue that, in contrast to YLW15, WL6 should have
essentially reached a slow rotation stage, with ``classical'' T Tauri (Class~II)
X-ray properties.

The outline of the paper is as follows.  We first concentrate on YLW15, studying
the implications of the quasi-periodic flares seen by {\sl ASCA} (\S2) in terms
of fast rotation of the central star:  we first establish the connection
between the flare periodicity and the rotation parameters of the star and the
accretion disk (\S2.1), then we study the long-term implications for the
magnetic braking of protostars (\S2.2) and for their angular momentum loss
(\S2.3); finally we point out some consequences on the X-ray flare energetics
(\S2.4).  We then turn to WL6, reexamining its light curve to conclude that it
is a slow rotator (\S3), and we derive the stellar parameters for YLW15 and WL6
(\S4).  From the comparison of the properties of these two Class~I protostars,
we examine briefly in \S5 the consequences of the fast rotation of protostars on
the magnetic braking time scale as a function of the stellar mass, and on the
accretion process; we also propose an explanation of the fact that few Class~I
protostars have been detected in X-rays so far, all in clusters; and
finally we discuss the main areas of uncertainty affecting these conclusions.
In the last section (\S6), we summarize the paper and conclude by a short
synthesis of the possible history of rotation and magnetic activity of
protostars.

\section{Implication of the quasi-periodic flares of YLW15}

	\subsection{The rotation connection}
	
		\subsubsection{X-ray periodicity observations of YLW15}

The {\sl ASCA} observations of YLW15 analyzed in Paper I show well-characterized
flaring with an approximate period $P_f \equiv 2\pi/\Omega_f \sim 20$\,hrs.
Fig.~\ref{YLW15} shows the X-ray light curve of YLW15 in the 2-10 keV domain.
At least the first flare can be well-modeled by the quasi-static radiative
cooling of a semi-circular magnetic loop-shaped tube (Van den Oord \& Mewe 1989)
of length $\sim 14 $\,R$_\odot$, i.e.  $\sim 4.5$\,R$_\odot$ in radius, and
aspect ratio $a =$ (cross-section diameter)/(loop length) $ \simeq 0.07$.  The
e-folding decay time of the flare is $\sim 30$\,ksec, i.e., less than half of
the flare period.  The peak luminosities are $L_{X,peak} \sim 2 \times
10^{32}$\,erg\,s$^{- 1}$ for the first flare, and $\approx 8-5 \times
10^{31}$\,erg\,s$^{-1}$ for the other two flares; the total energies released
are respectively $E_{tot} \sim 6.5 \times 10^{36}$\,erg for the first flare, and
$\approx 3 \times 10^{36}$\,erg for the second and third flare.  The values
derived in Paper I for the plasma density ($n_e \sim 5 \times 10^{10}$
cm$^{-3}$) and the equipartition magnetic field of each flare ($B_{eq} \sim 100
- 150$ G) are comparable to those of solar coronal plasmas, which justifies here
again, as in all YSOs, the use of the solar analogy as a basis to analyze the
implications of the triple flare of YLW15.  (For reference, see the review by
Feigelson \& Montmerle 1999:  FM.)

\vspace{0.25cm}
\centerline{\hbox{\psfig{file=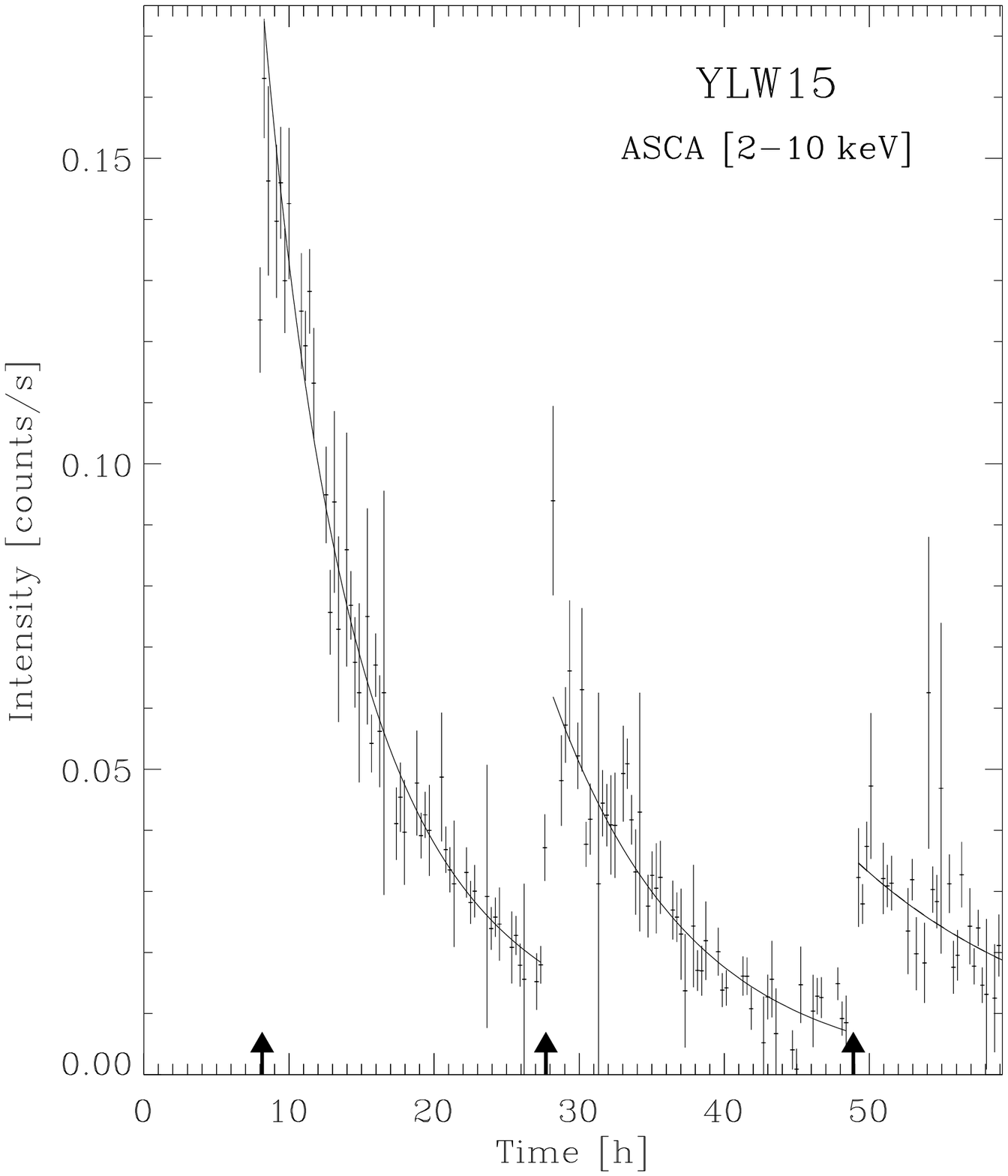,width=0.45\textwidth}}}
\vspace{-0.5cm}
\figcaption{\small Light curve of YLW15 obtained with {\sl ASCA} (Tsuboi et al. 1999: Paper I). 
The data points of the first flare are fitted with a quasi-static radiative 
cooling model, and those of the other two flares are fitted by exponentials
with radiative cooling timescales (see Paper I for details). The arrows
point at the start of the flares, with a characteristic interval $\sim 20$h.\label{YLW15}}
\vspace{0.25cm}

Now the {\sl ASCA} observation of Paper I could be considered as a random event.
However, there is some evidence that the same $\sim 20$\,h period was already
present in {\sl Tenma} observations over a decade ago (Koyama 1987).  {\sl
Tenma} was a wide-field, non-imaging satellite, sensitive to hard X-rays like
{\sl ASCA}.  The observation covered the whole $\rho$ Oph cloud.  The derived
temperature ($kT \sim 4$\,keV) and X-ray luminosity ($L_X[1.5-10\,{\rm keV}]
\sim 1.8 \times 10^{32}$\,erg\,s$^{-1}$) are comparable with those of the triple
flare observed with {\sl ASCA} ($kT \sim 2-6$\,keV, and $L_X[2-10\,{\rm keV}]
\sim 11- 2 \times 10^{31}$\,erg\,s$^{-1}$).  These properties were from the
start puzzlingly different from those derived from {\sl Einstein} images of the
same region, which showed only T Tauri stars with $kT \sim 1$ keV and $L_X(0.5
- 4.5 {\rm keV}) \sim 2-20 \times 10^{30}$ erg s$^{-1}$ (Montmerle et al.
1983).  Nowadays, it is well known from {\sl ASCA} images showing both
protostars and T Tauri stars in the field-of-view that protostars are
preferentially detected in the hard X-ray band ($\simgt$ a few keV), in contrast
with the dominant population of T Tauri stars, which generally emit softer
X-rays and are less luminous (e.g., Fig.~1 of Paper I, and Figs.~1$a$ and 1$b$
of Koyama et al.  1996).  In retrospect, this {\sl Tenma} result is therefore
totally consistent with the interpretation that YLW15 dominated the X-ray flux
of all YSOs in the field-of-view at the time of the observation.

This suggests that the characteristic timescale of $\sim 20$\,h is a long-lived
feature, directly related to an intrinsic property of YLW15, and not related to
flare events {\it a priori} randomly distributed in time.  The simplest
explanation is that {\it the X-ray flare period must be closely related to the
stellar rotation period.}

	\subsubsection{Star-disk magnetic interactions}

The analysis of the X-ray properties of the YLW15 triple flare
presented in Paper I indicates that we are certainly in the framework of
magnetically induced X-ray activity classically invoked for this type of object
(see, e.g., FM).  However, the X-ray periodicity underlying the flares is a
novel feature which cannot be easily explained in terms of stellar activity
alone.  Given the existence of a previously observed X-ray ``superflare'' on the
same object, with the unavoidable necessity of invoking some form of star-disk
magnetic interaction as a corollary (Grosso et al.  1997), we also adopt this
framework to interpret the origin of the triple flare.

Calculations of magnetic configurations resulting from star-disk magnetic
shearing have recently been performed in the context of the origin of jets and
outflows from young stellar objects (YSOs).  These calculations have shown that
an initially poloidal field line, anchored both in the star and in the disk, is
twisted and inflated when the footpoints are not rotating synchronously.  As a
result, an increasingly large toroidal component develops (see, e.g, the helical
configuration of Lovelace, Romanova, \& Bisnovatyi-Kogan 1995).  This
configuration allows antiparallel segments of the field lines to be brought in
contact and give rise to reconnection.  Theoretically, this reconnection can
take place within the same field line, as in the solar case (giving rise to
the``helmet'' configuration, e.g., Hayashi, Shibata, \& Matsumoto 1996), or
between different field lines, as may happen for accreting neutron stars (Aly \&
Kuijpers 1990).  So far, the field topology has been calculated only in 2D and
2.5D (see the numerical simulations by Hayashi, Shibata, \& Matsumoto 1996;
Goodson, Winglee, \& Boehm 1997; Hirose et al.  1997; Miller \& Stone 1998), and
in 3D assuming axisymmetry (Fendt \& Elstner 1999), but the Sun gives many
observational examples of distorted topologies giving rise to reconnection
(e.g., Innes et al.  1997).  Additional mechanisms include resistive tearing
instabilities in YSO magnetosphere-disk interactions (Birk 1998), and, as was
studied numerically in the solar case by Amari et al.  (1996), the twisting of a
magnetic tube rooted in counter-rotating convective cells in the photosphere.
Obviously, there is no shortage of possible reconnection scenarios.

\vspace{0.25cm}
\centerline{\hbox{\psfig{file=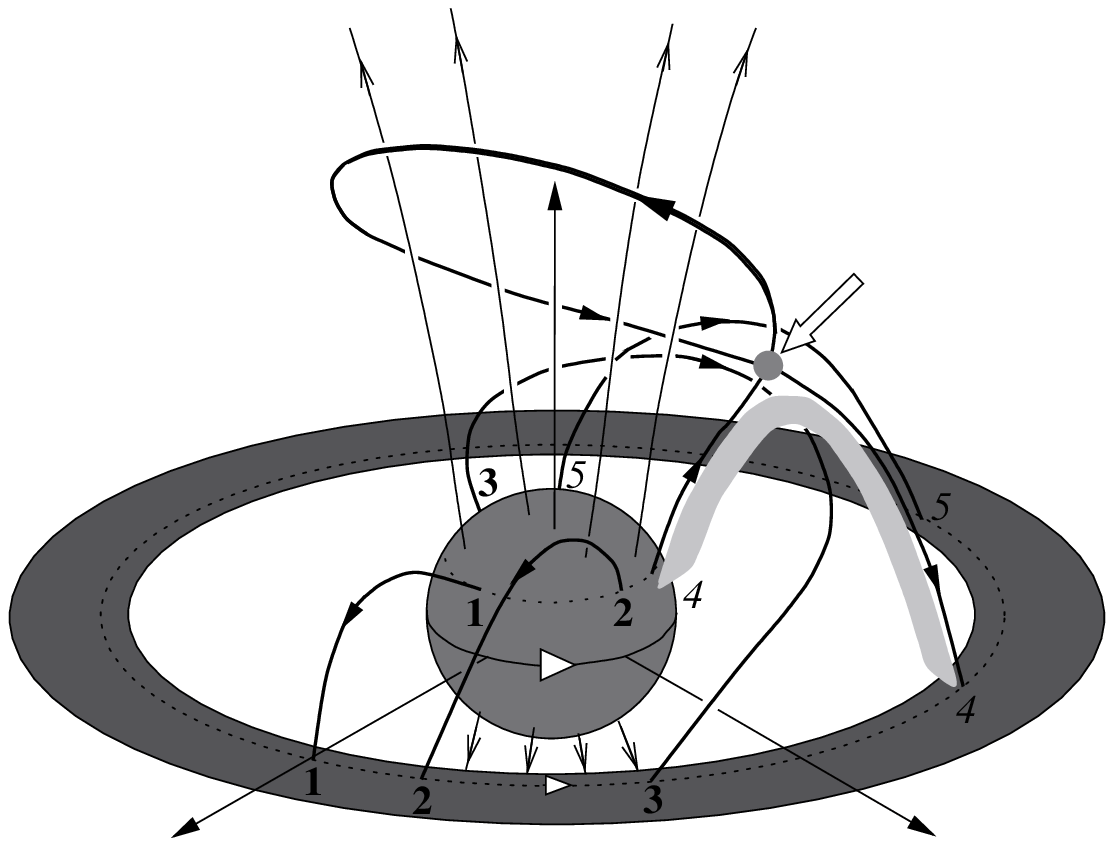,width=0.45\textwidth}}}
\figcaption{\small Sketch of the plausible evolution of a magnetic field configuration
leading to reconnection and flaring for the quasi-periodic light curve of YLW15.
This is just one of the possible configurations (see text, \S 2.1.2 for
discussion).  The open lines are other, unperturbed field lines.  This
figure schematically describes five steps of the evolution of a field line
assumed to initially connect the star of period $P_\star$ and the inner disk of
period $P_{disk}$ (step 1).  For illustrative purposes, we have chosen $P_\star
= P_{disk}/4$, thus $P_{beat} \equiv \mid 1/P_\star - 1/P_{disk} \mid^{-1} =
4P_\star/3$:  at steps 4 and 5 (in italics), the star has rotated more than one
period.  Since the star rotates faster than the inner disk, at steps 2 and 3 the
magnetic field is sheared between the star and the disk, and the loop
progressively inflates and surrounds ordered open stellar magnetic field lines.
At step 4, the stellar footpoint has moved one beat period, so the initial field
line comes into contact with itself (open arrow) and reconnects.  The magnetic
energy thus liberated heats the plasma and ignites a flare.  (The shaded
loop-like ``brush stroke'' schematically underlines the loop remaining after
reconnection derived in Paper I, which we use for calculations in the present
paper.) Step 5 illustates the post-reconnection configuration, in which the
plasma is radiatively cooling by X-ray bremsstrahlung and line emission.  The
scenario can then repeat itself if the magnetic field of step 4 reverts to the
initial configuration after the reconnection (after step 5); in that case,
successive reconnections and flaring may continue periodically for some time.\label{schema}}
\vspace{0.25cm}

Fig.~\ref{schema} illustrates one plausible scenario for the evolution of the
configuration of magnetic field lines connecting the star and the accretion
disk.  Such an evolution leads to growth of the magnetic energy stored in the
field lines at the expense of the star-disk mechanical energy, followed by local
reconnection resulting in fast heating of the plasma and flaring.  The figure
schematically describes five steps of the evolution of a field line (or more
realistically, as we have seen in Paper I, a thin tube) assumed to initially be
anchored in the star and in the disk.  The caption gives some details for each
step.  In brief, since the star rotates faster than the disk (this is
demonstrated below), the magnetic field is sheared between the star and the
disk, and the trailing loop progressively inflates and surrounds ordered open
stellar magnetic field lines.  The magnetic energy liberated when reconnection
occurs heats the plasma and ignites a flare.  The scenario can then repeat
itself if the anchored part of the magnetic field reverts to the initial
configuration after the reconnection; in that case, successive reconnections and
flaring may continue periodically, at least for some time.

We have chosen this type of scenario because of the fact that in the
quasi-periodic flares of YLW15 observed by {\sl ASCA}, the radii of the plasma
confining loops do not change much in the three successive flares ($R_{loop}
\sim 4.5 R_\odot$), and radiative cooling models give good fits to their light
curves.  As discussed in detail in Paper I, these observations are consistent
with the conclusion that {\it some basic magnetic structure, approximated by a
semi-circular loop, has been reconstituted three times after reconnection, in a
periodic fashion}.  This is schematically illustrated in Fig.~2 in the form of a
shaded loop-like ``brush stroke''.

	\subsubsection{Determination of the star and disk rotation parameters}

Taking this interpretation as a starting point, we are now in a position to
derive the rotation parameters of the star and the disk in three steps:  ($i$)
investigate the relative kinematics of the star and the inner disk, where the
loop footpoints are assumed to be anchored, ($ii$) use solar plasma physics to
set constraints on star-disk magnetic field shearing and flare ignition, and
($iii$) combine the results.

Let $A$ and $B$ be the magnetic footpoints separated by a distance $2R_{loop}$,
respectively located on the star (radius $R_\star$) and the disk (radius $r_D
\approx R_\star + 2 R_{loop}$):  $A$ rotates at some unknown stellar period
$P_\star$, and the Keplerian rotation period of the disk at $B$ is $P_D \equiv
2\pi/\Omega_D$.  We here adopt the view that the field lines and matter are
strongly coupled:  this is justified by the fact that a high ionization rate
must be present in the disk, precisely because of X-ray irradiation (see
Glassgold, Najita, \& Igea 1997). In our scenario, the observed flare period
$P_f$ is therefore some fraction $\alpha$ of the ({\it a priori} unknown) beat
period $P_{beat}$ between $A$ and $B$:

\begin{equation}  P_f = \alpha P_{beat}
\label{eq:def_alpha}
\end{equation}

\noindent
with

\begin{equation} 1/P_{beat} \equiv \mid 1/P_D \pm 1/P_\star \mid.
\end{equation}

Adopting the natural hypothesis that the star and the disk
rotate in the same direction, and expressed in terms of positive periods,
this equation can be rewritten as:

\begin{equation}
P_\star = \left\{\begin{array}{llr}
			1/(1/P_D+1/P_{beat}) & \mbox{if $P_\star \le P_D$,} \\ 
			1/(1/P_D-1/P_{beat}) & \mbox{if $P_\star \ge P_D$.}
		\end{array}  \right. 
\label{eq:periode_etoile}
\end{equation}

In our case, for a given stellar mass and radius, $P_D$ is a function of the
star-disk distance $r_D$, itself ultimately fixed by the flaring loop size.
Figure~\ref{periode_disque_etoile} schematically shows $P_\star$ as a function
of $P_D$ (Eq.~3) in the form of two half hyperbolae, with asymptotes at
$P_{\star} = P_{beat}$ and $P_D = P_{beat}$.  For $P_D < P_{beat}$, there are
two solutions, above and below the $P_\star = P_D$ bisector:  one corresponds to
$P_\star < P_D$ (the star rotates faster than the disk), the other to $P_\star >
P_D$ (the disk rotates faster than the star).  \footnote {Such a situation had
been invoked in the past to explain spectral features of classical T Tauri
stars, such as emission lines and veiling, in terms of a boundary layer between
a slowly rotating star and an accretion disk rotating at (fast) Keplerian
velocities in contact with the star (see Bertout, Basri \& Bouvier 1988).}  For
$P_D > P_{beat}$, there is only one solution, with $P_\star < P_D$.  These
solutions may be interpreted as follows:  for small values of the disk period
(i.e., close to the star, or $r_D$ small), the disk and the star have comparable
periods, and the flare loop can be either trailing or leading.  When the
difference between the star and disk periods becomes large enough ($r_D$ large),
the star must rotate much faster than the disk, and only the trailing solution
remains.  The limiting cases of the asymptotes are when the disk is at infinity
(then $P_\star = P_{beat}$), or when the star does not rotate (then $P_{disk} =
P_{beat}$).  The question now becomes:  in which regime are we ?

\vspace{0.25cm}
\centerline{\hbox{\psfig{file=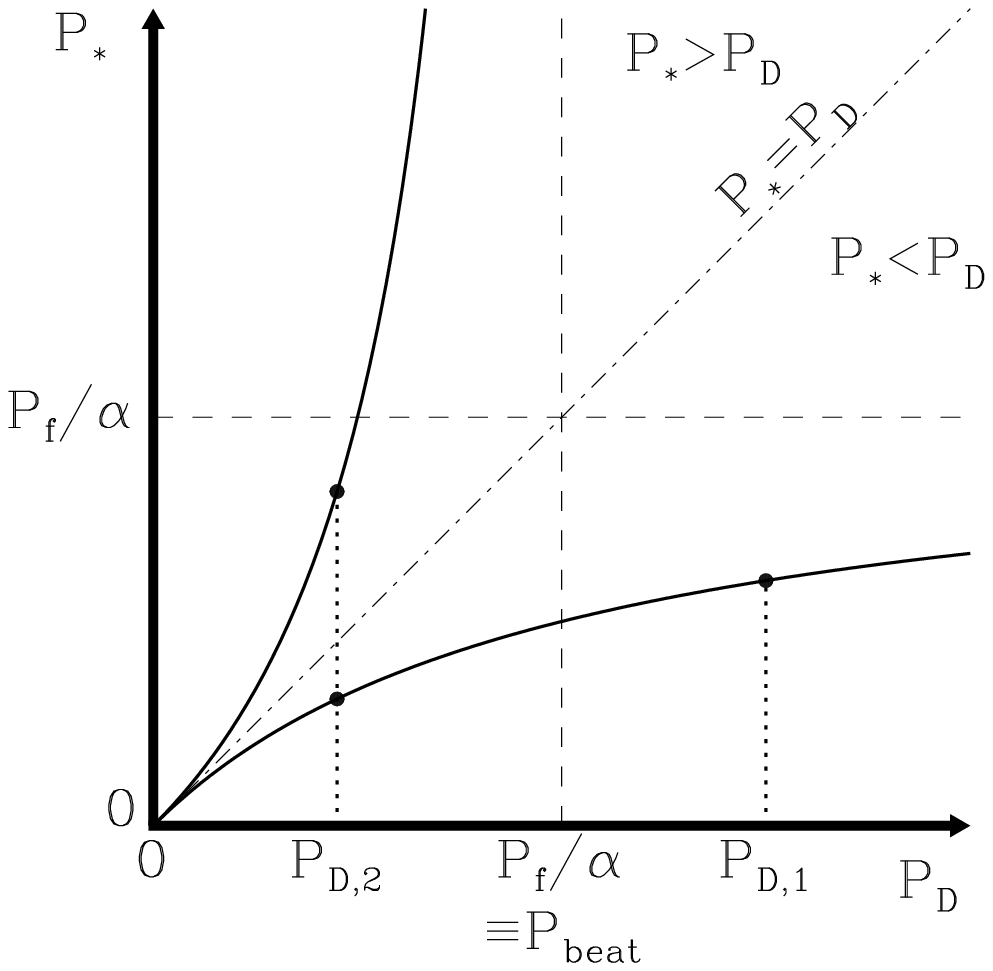,width=0.4\textwidth}}}
\figcaption{\small Stellar period ($P_{\star}$) as a function of disk period ($P_D$).
The possible solutions are in the form of two half hyperbolae, with asymptotes
at $P_{\star} = P_{beat}$ and $P_D = P_{beat}$.  For $P_D = P_{D,2} < P_{beat}$,
there are two solutions:  one corresponds to $P_\star < P_{D,2}$ (the star
rotates faster than the disk), the other to $P_\star > P_{D,2}$ (the disk
rotates faster than the star).  For $P_D = P_{D,1} > P_{beat}$, there is only
one solution, with $P_\star < P_{D,1}$. \label{periode_disque_etoile}}
\vspace{0.25cm}

For a star of mass $M_\star$, the disk Keplerian rotation period $P_D$ at $B$
is:  
\begin{equation} P_D = 2\pi r_D^{3/2} (GM_\star)^{-1/2},
\label{eq:periode_disque} 
\end{equation} 

\noindent recalling that $r_D \approx R_\star+ 2R_{loop}$.  We will use below
(\S 4) a ``mass-radius'' relation between $M_\star$ and $R_\star$, and find that
the bolometric luminosity constrains the stellar mass to be $M_\star \simlt
2.2$\,M$_\odot$ (hence $R_\star \sim 4.2 R_\odot$).  Thus, $r_D$ being a
function of $R_\star$, we find via Eq.~(\ref{eq:periode_disque}) $P_D$ as a
function of $M_\star$ only:  this is shown on Fig.~\ref{periode_disque_masse}.
Fixing $M_\star = 2.2$\,M$_\odot$ yields $P_D \equiv P_{D,\star} = 3.8$ days.
In analogy with Eq.~(\ref{eq:def_alpha}), we then define $\alpha_\star \equiv
P_f/P_{D,\star} = 0.22$.  Depending on whether $\alpha$ in
Eq.~(\ref{eq:def_alpha}) is smaller (resp.  larger) than $\alpha_\star$, there
will be two (resp.  one) solution to Eq.~(3).

We can now use results obtained in the framework of solar plasma physics.  Amari
et al.  (1996) have made a 3D numerical simulation of the MHD evolution of a
tubular magnetic loop anchored to footpoints twisted by some underlying
movements (such as caused by convection on the solar surface).  Their results
demonstrate that it takes almost one twisting period $\tau_{twist}$ between the
footpoints for field lines to begin to open:  Fig.~4 of Amari et al.  shows that
the opening process is very slow at first, starting at $\approx 0.7
\tau_{twist}$, then proceeds very fast.  The field lines are completely open at
$\sim 0.8 \tau_{twist}$.  Since the opening of field lines is a prerequisite for
reconnection to occur at some later time $\tau_{rec}$, we rewrite this result in
the form $\tau_{rec} = \alpha_{rec} \tau_{twist}$, with the ``reconnection
parameter'' $\alpha_{rec} \sim 0.8 - 1$:  by doing so, we allow for a small
random flare-to-flare change ($\leq 20\%$) in the moment when reconnection
actually occurs, after the start of the opening of the field lines.

In YLW15, the X-ray flaring is produced by a star-disk shearing process having a
characteristic time scale which is the relative period between the star and the
disk, i.e., $P_{beat}$;  It is clear that the corresponding magnetic
configuration (Fig.~\ref{schema}) is not the configuration of Amari et al.

\centerline{\hbox{\psfig{file=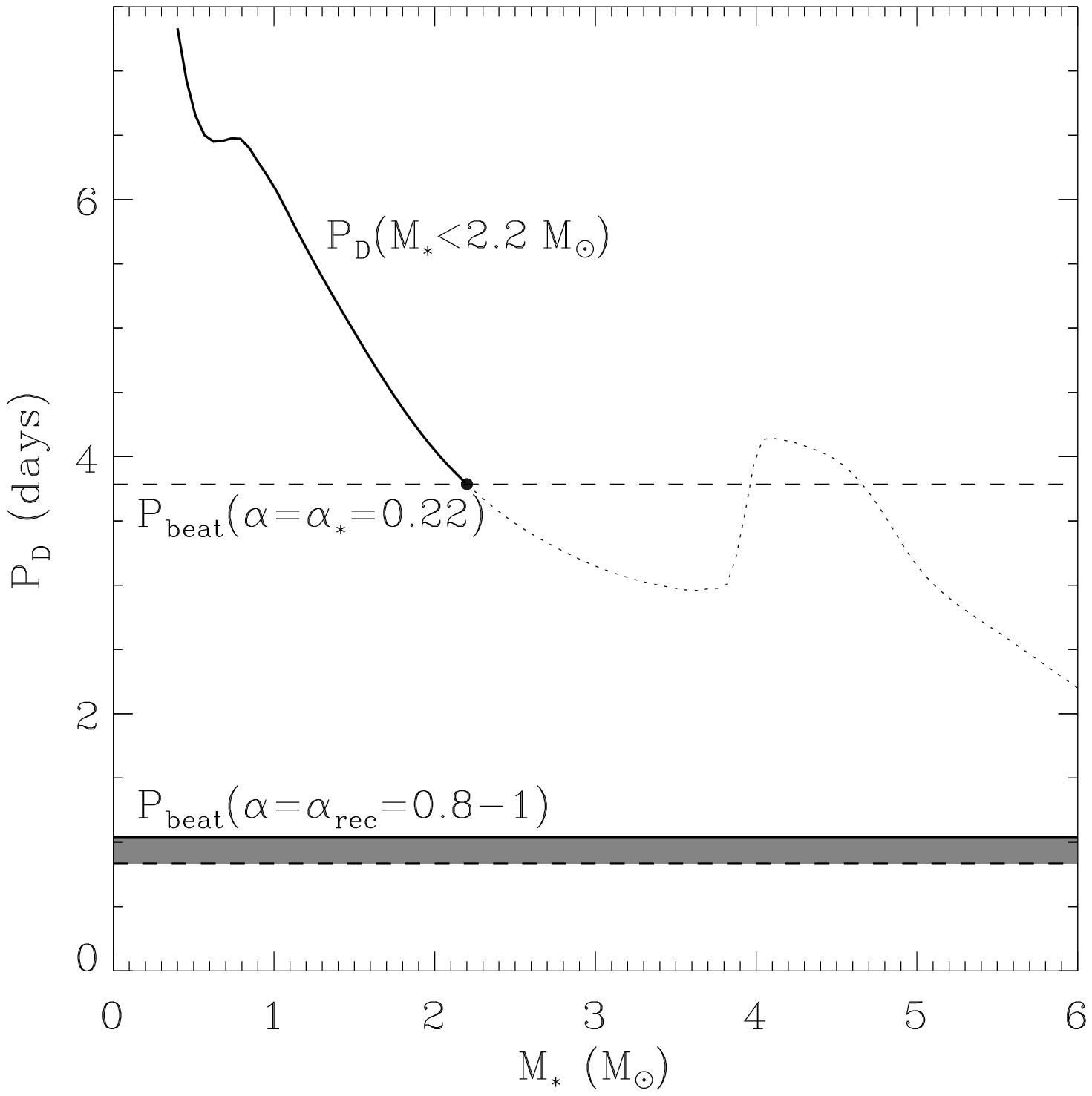,width=0.45\textwidth}}}
\figcaption{\small Comparison between the estimated disk period ($P_D$)
and the star-disk beat period ($P_{beat}$).  $P_D$ is a function of the stellar
mass ($M_\star$), the stellar radius ($R_\star$), and the flaring loop size.
Using a ``mass-radius'' relation, and keeping the flaring loop size fixed
(estimated in Paper I), $P_D$ becomes a function of $M_\star$ only (dotted
line).  For YLW15, the bolometric luminosity gives an upper limit to the stellar
mass of $M_\star \sim 2.2\,M_\odot$ (bold line; see text, \S 4), from which we
obtain $P_D \equiv P_{D,\star} = 3.8$ days.  $P_{beat}$ is proportional to the
flare period $P_f$, via the parameter $\alpha$ ($P_{beat}=P_f/\alpha$).  For
$\alpha=\alpha_\star \equiv P_f/P_{D,\star}=0.22$, $P_{beat}=P_{D,\star}$, and
there are two solutions for the stellar period:  either the star does not
rotate, or the star rotates with $P_\star=P_{beat}/2$.  On the other hand, solar
plasma physics tell us (see discussion in the text, \S 2.1.3) that the
``reconnection parameter'' $\alpha_{rec}$, characterizing the moment when flare
ignition occurs after magnetic stresses are applied, is $\alpha_{rec} = 0.8 -
1$.  The corresponding beat periods are respectively indicated by a continuous
and a dashed horizontal lines, separated by a grey strip.  Since $\alpha_{rec} >
\alpha_\star$, there is finally only one solution, in which the star rotates
faster than the disk.\label{periode_disque_masse}}
\vspace{0.25cm}

However, their results suggest that reconnection happens at the very end of the
stress build-up process which has a characterictic time scale:  $P_{beat}$ in
our case, $\tau_{twist}$ in the case of Amari et al.  It is thus reasonable to
{\it identify} $P_{beat}$ with $\tau_{twist}$ and $P_f$ with $\tau_{rec}$. 
Then Eq.~(\ref{eq:def_alpha}) can be understood as expressing the time lag separating
the start of shearing process between the star and the disk, and the moment of
flare ignition.  Therefore, the parameter to be compared to $\alpha_\star$ is
$\alpha_{rec} \equiv P_f/P_{beat}$.  With $\alpha_\star = 0.22$, we find that
$\alpha_{rec} > \alpha_\star$.

We conclude that we are in the regime where there is only one solution, that of
a star rotating faster than the disk.  As illustrated on
Fig.~\ref{periode_disque_masse}, $\alpha_{rec} > \alpha_\star$
throughout the allowed stellar mass range, and the difference between
$\alpha_{rec}$ and $\alpha_\star$ is so large that this conclusion weakly
depends on the actual value of $\alpha_{rec}$, although the
resulting stellar period $P_{\star}$ admittedly depends on it (via $P_{beat}$).
For $\alpha_{rec} = 0.8 - 1$ as argued above, we find from
Eq.~(\ref{eq:periode_disque}) $P_{\star}=20-16$\,h.  In other words, {\it the
central star of YLW15 must be a fast rotator}, having a rotation period shorter
than the observed X-ray flare period, far from corotation with the inner part of
the disk, which rotates much more slowly.

\vspace{1cm}
	\subsection{Magnetic braking}

The situation in YLW15 is therefore {\it not} the ``classical'' quasi
steady-state situation argued to hold in protostars (e.g., Shu et al.  1997) or
in T Tauri stars (K\"onigl 1991), but rather {\it precedes it}.  The framework
is reminiscent of disk-fed compact X-ray binaries, where a magnetized neutron
star accretes material from a companion via a disk (see, e.g., Aly 1986 for an
extensive review).

In the steady-state approximation, the magnetic field rotates as a solid body
tied to the star, basically up to the Alfv\'en radius $r_A$ where the ram
pressure from the disk accretion flow (at a rate $\dot{M}_{acc}$) is balanced by
the magnetic pressure.  The basic idea, first proposed by Ghosh \& Lamb (1978)
for a neutron star having a dipolar magnetic field and surrounded by an
infinitely thin accretion disk, is that the pressure of the inwards-moving
material keeps field lines open down to a radius comparable to $r_A$.
Conversely, the magnetic pressure inside this radius is strong enough to prevent
the accreting material in the disk plane from spiralling in towards the star,
but rather lifts it into free fall along the field lines.  This concept of
``magnetospheric accretion'' in classical T Tauri stars (review by Shu 2000) has
now been successfully modeled to interpret line emission features (e.g., Edwards
et al.  1994; Muzerolle, Hartmann, \& Calvet 1998), photometric variations
(Bouvier et al.  1999), as well as the broadening of some absorption lines
(Johns-Krull, Valenti, \& Koresko 1999).  The border region between the closed
magnetosphere and the disk is in reality poorly known, with differential
rotation and complex exchanges of angular momentum between the star and the
disk, and depending strongly on the assumed magnetic diffusivity of the
accretion disk, i.e., on the way the stellar magnetic field will actually
penetrate the disk material.

Pursuing this issue further is beyond the scope of this paper.  We will now
simply follow the classic concept of ``magnetic braking'' to describe the {\it
long-term evolution} of the star-disk system in the idealized situation where
the inner border of the disk is located near the Alfv\'en radius, asymptotically
ending up in ``magnetic locking'', where the star and the inner disk corotate
via a rigid magnetosphere (see, e.g., the details of the model by Shu et al.
1997) as is presumably the case in classical T Tauri stars (see below, \S2.4).

In the above framework, the characteristic magnetic braking time scale
$t_{br}$ is obtained by equating the angular momentum variation and the
magnetic torque (e.g., K\"onigl 1991, and references therein):

\begin{equation} d(J_\star \Omega_\star)/dt \sim \mu_\star^2/r_A^3, 
\label{moment_cinetique}
\end{equation}

\noindent 
with $r_A = \beta \mu_\star^{4/7}(2GM_\star)^{- 1/7}\dot{M}_{acc}^{-2/7}$ 
for a dipolar magnetic field. 

The factor $\beta$ takes into account the geometry of the accretion:
$\beta = 1$ for spherical accretion, and $\beta = 0.52$ for accretion from an
infinitely thin disk (Ghosh \& Lamb 1979).  In the case of protostars and T
Tauri stars, however, the disk is rather geometrically thick (see the ``flared
disk'' models introduced by Kenyon \& Hartmann 1987, and the state-of-the-art
observational constraints by Bell et al.  1997), and protostars also have
circumstellar envelopes and outflows.  Any distant coupling between the magnetic
field and matter will make $\beta$ tend towards 1:  in fact, in some accreting
neutron stars $\beta$ has actually been {\it measured} to be $\sim 1$ (Wang
1996).  As a result, for protostars it is likely that $\beta \simlt 1$.

From Eq.~(\ref{moment_cinetique}), the braking time $t_{br}$ can then be expressed as:

\begin{equation} t_{br} \sim \beta^3 J_\star \Omega_\star \mu_\star^{-2/7} 
(2G M_\star)^{-3/7} \dot{M}_{acc}^{-6/7}, \end{equation}

\noindent where $J_\star = \gamma(M_\star) M_\star R_\star^2$ and 
$\mu_\star = B_\star R_\star^3$ are respectively the moment of 
inertia and the magnetic moment of a star of mass $M_\star$, radius $R_\star$,
and surface magnetic field $B_\star$. For young stars at the 
``birthline'', the coefficient $\gamma(M_\star) \simeq 1$ for masses in 
the range $M_\star \sim 1 - 3 $\,M$_\odot$ (Palla 1999, priv. comm.; 
from the models of Palla \& Stahler 1992), so that

\begin{equation}
t_{br} \sim (2G)^{-3/7} \beta^3 \Omega_\star B_\star^{-2/7} M_\star^{4/7} R_\star^{8/7}
\dot{M}_{acc}^{-6/7}.
\end{equation}

\noindent
If the initial angular velocity is the break-up angular velocity
 $\Omega_{\star,0} = (GM_\star)^{1/2} R_\star^{-3/2}$,
the corresponding breaking time $t_{br,0}$ is

\begin{equation}
t_{br,0}  \sim  2^{-3/7} G^{1/14} \beta^3 B_\star^{-2/7} M_\star^{15/14} R_\star^{-5/14}
\dot{M}_{acc}^{-6/7}.
\end{equation}

\noindent Therefore $t_{br,0}$ is almost proportional to the stellar mass,
almost inversely proportional to the accretion rate, but depends comparatively
weakly on $B_\star$ and $R_\star$.

For YLW15, the X-ray flare modelling gave a uniform equipartition magnetic field
strength $B_{eq} \sim 150$\,G.  It is not completely clear to what extent the
magnetic loop deduced from this flare could be related to some underlying
large-scale magnetic structure.  Assuming for simplicity that this structure is
a dipolar magnetosphere, $B_{eq}$ would be a measure of the value at the top of
a dipolar loop, in which case the photospheric flux $B_\star$ would be more like
$\approx 1$\,kG.  In view of the uncertainties, we take $B_\star \sim 0.1 -
1$\,kG.  As regards the relevant accretion rate, we will argue below (\S 4) that
it is {\it not} the accretion rate necessary to build up the star ($\approx$
1\,M$_\odot$ in $\approx 10^5$\,yrs, i.e., $\dot{M}_{acc,\star} \sim
10^{-5}$\,M$_\odot$ yr$^{-1}$; see Palla \& Stahler 1992), which corresponds to
the Class 0 protostar phase, but rather the presently observed Class~I disk
accretion rate deduced from the outflow or the bolometric luminosity, which is
about 10 times smaller.  Thus within a factor of $< 2$ if $\beta$ lies
between $\beta \sim 0.75$ (see below, \S5.1) and $\beta =
1$, we express $t_{br,0}$ numerically as

\begin{eqnarray} 
t_{br,0} & \sim & 3.3 - 7 \times 10^5\,yrs \times \left(\frac{B_\star}{1000\,\rm G}\right)^{-2/7} \left(\frac{R_\star}{R_\odot}\right)^{-5/14} \nonumber\\
	 &      & \times \left(\frac{M_\star}{M_\odot}\right)^{15/14} \left(\frac{\dot{M}_{acc}}{10^{-6}\,M_\odot~{\rm yr}^{-1}}\right)^{-6/7}.
\label{eq:braking_time}
\end{eqnarray}

Taking $M_\star \approx 2$\,M$_\odot$, and $R_\star \approx 4$\,R$_\odot$ from
the models by Palla \& Stahler (1992) for the central star of YLW15 (see \S4),
and the above values for the other parameters, we find $t_{br,0} \sim 4.3 - 17
\times 10^5$\,yrs for $B_\star \sim 1 - 0.1$\,kG and $\beta \sim 0.8 - 1$.  With
a build-up accretion rate $\dot{M}_{acc,\star} \sim 10^{-5}$\,M$_\odot$
yr$^{-1}$, the age of YLW15 is $\sim 2 \times 10^5$\,yrs, i.e., smaller than
$t_{br,0}$.  This suggests that the central star of YLW15 is still decelerating,
and has not yet been brought to a steady-state corotation with the inner
part of the accretion disk.  (We neglect here any possible deceleration
having taken place during the earlier Class 0 stage, which, lasting $\sim 10^4$
yrs [AWB], is very short.  In addition, at this stage there is no ``real'' star,
but mainly an infalling rotating envelope [see, e.g., Terebey, Shu, \& Cassen
1984], so that angular momentum may actually be {\it gained} during this phase;
see also the discussion in \S 6.  In addition, we have no idea of the magnetic
field origin, configuration, or intensity at this stage, so that Eq.~[9] cannot
be applied anyway.)

As a final consistency check between the long-term evolution given by $t_{br}$
and the short-term magnetic shearing leading to reconnection, we must verify
that the Alfv\'en radius $r_A$ is not larger than the disk radius at footpoint
$B$, $r_D$ $(\sim 3.5 R_\star)$.
We find numerically:

\begin{eqnarray} 
r_A 	& = & 1.1 R_\odot \times \beta \left(\frac{B_\star}{1000\,G}\right)^{-4/7} \left(\frac{R_\star}{R_\odot}\right)^{-12/7} \nonumber\\
	&   & \times \left(\frac{M_\star}{M_\odot}\right)^{-1/7} \left(\frac{\dot{M}_{acc}}{10^{-6}\,M_\odot {\rm yr}^{-1}}\right)^{-2/7}.
\label{rayon_alfven}
\end{eqnarray}

We thus have $r_A \simlt r_D$ for $B_\star \simlt 950 - 1900$ G ($\beta = 0.5
-1$), which is compatible with the X-ray derived values of $B_\star$.

	\subsection{A collimated coronal wind ?}

In our case, as a consequence of the assumed X-ray--induced strong magnetic
coupling between the star and the disk, there is no rigid magnetosphere, but
rather a succession of readjustments via shearing and reconnection.  There will
be energy losses and momentum exchanges between the star and the disk, leading
to ``magnetic braking''.  Energetic X-ray flares can be taken as a tracer of the
energy lost in radiative form during this braking (see below).  In solar flares
and other manifestations of coronal activity, material is ejected after
reconnection events by a variety of processes.  So we expect the star
to eject material in some form of {\it coronal wind}.  We can obtain
an order-of-magnitude estimate of the corresponding mass-loss rate during a
flare such as the ones discussed here by assuming that this wind comes from the
plasma ``rings'' left after reconnection, in a fashion analogous to solar
coronal mass ejections.  The mass of this detached part, approximated by a ring
of radius $R_{ring}$ and cross-section diameter $2\pi R_{ring} \times a$, will
be

\begin{equation}
m_{ej} \approx 2\pi
R_{ring} \times \pi^3 a^2 R_{ring}^2 \times n_e m_p
= 2 \pi^4 a^2 R_{ring}^3 n_e m_p.
\end{equation}

Taking Fig.~2 at face value, we adopt $R_{ring} \approx R_\star + R_{loop}
\approx 8.7 R_\odot$.  Also (Paper I), the aspect ratio $a = 0.07$, and the
plasma density $n_e = 5 \times 10^{10}$ cm$^{-3}$.  We thus find $m_{ej} = 1.8
\times 10^{22}$ g per flare event.  Taking this mass to be ejected every $P_f
\sim 20$ h, the resulting mass-loss rate is $\dot{M}_{ej} = m_{ej}/P_f \sim 3.9
\times 10^{-9} M_\odot$ yr$^{-1}$.  However, it is important here to realize
that such a coronal wind {\it should not be confused} with the outflows
currently associated with protostars, which have a mass-loss rate several orders
of magnitude larger ($\dot{M}_{out} \approx 10^{-5} M_\odot$ yr$^{-1}$ for Class
I protostars, see Bontemps et al.  1996).  In most models, such outflows result
from a (cold) {\it disk} wind, completely distinct from what happens in the
vicinity of the central star inside the Alfv\'en radius (e.g., Camenzind 1997,
K\"onigl \& Pudritz 2000).

If this order of magnitude for $\dot{M}_{ej}$ is correct, then we should worry
about a number of consequences.  

($i$) Being free from star and disk material, the magnetic ring will not confine
the plasma anymore, which will therefore expand and cool via radiative and
adiabatic losses; based on the analogy with the solar wind, heating (for
instance by Alfv\'en waves) may also happen.  We can speculate that the
equilibrium temperature will be a ``typical'' coronal temperature $T_e \sim
10^6$ K.  As a result, the plasma will radiate in soft X-rays ($\simlt 0.1$
keV), which are totally absorbed by the intervening material, hence be
undetectable.

($ii$) However, {\it this plasma will also radiate in the centimeter radio
range}, which suffers no extinction by the surrounding matter unless this matter
is fully ionized.  We have mentioned in the Introduction (\S1) that YLW15 is the
seat of the highest radio emission, likely thermal, of all the VLA sources in
the $\rho$ Oph core F, with a total flux density $F_{5GHz} = 3.3$ mJy.  It is
thus tempting to compare this value with the free-free emission from the
putative coronal wind of YLW15.

The scenario illustrated in Fig.~2 suggests that we are dealing with a bipolar
geometry.  Then, for a standard ionized wind of temperature $T_e$, contained in
a cone of aperture $\theta_0$ inclined at an angle $i$ to the observer, and
flowing at velocity $v_w$, the mass-loss rate $\dot{M}_{ej}$ corresponding to a
free-free radio flux density $F_\nu$ at frequency $\nu$ is given by (from
Reynolds 1986, Andr\'e 1987; for $\theta_0 \simlt 0.5$):

\begin{eqnarray}
\dot{M}_{ej} 	& = & 5.2 \times 10^{-9}M_\odot {\rm yr}^{-1}~ \theta_0^{\,3/4} (\sin{i})^{-1/4} \left(\frac{F_\nu}{\rm mJy}\right)^{3/4} \nonumber\\
		&   & \times \left(\frac{v_w}{\rm 100\, km/s}\right) \left(\frac{\nu}{\rm 5\, GHz}\right)^{-0.45} \left(\frac{T_e}{10^4\, {\rm K}}\right)^{-0.075} \nonumber\\
		&   & \times \left(\frac{d}{160\, {\rm pc}}\right)^{1.5}.
\end{eqnarray}

Strictly speaking, this relation is valid for $kT_e \simlt 0.05$ keV; above this
value, quantum effects become important, but the temperature dependence of the
emissivity retains a logarithmic term (Oster 1961), so that for simplicity we
will keep the same equation for $kT_e$ up to $\sim 0.1-1$ keV.  In Eq.~(12)
above, the plasma temperature plays in practice a very small role, which is
fortunate because we do not know the actual temperature structure of the coronal
wind.  For YLW15, we can take $\sin{i} = 1$, $v_w \approx 300$ km s$^{-1}$ (escape
velocity from the surface of the central star), and we adopt $T_e = 10^6$ K.  We
know all the other parameters, except the opening angle $\theta_0$.  Gathering
numbers, we find:

\begin{equation}
\theta_0 ({\rm YLW15}) = 0.012~\left(\frac{\dot{M}_{ej}}
{10^{-9}\,M_\odot {\rm yr}^{-1}}\right)^{4/3},
\end{equation}

\noindent
which, taking the value of $\dot{M}_{ej}$ we have derived, gives $\theta_0
= 0.08$, or about 5$^{\circ}$. This good collimation (in reality a conical
approximation to the actual wind flow) suggests that such a coronal wind
could be related to the jets frequently seen in the close vicinity of YSOs
(e.g., Rodr\'{\i}guez 1997).

($iii$) Making VLBI observations of YLW15, Andr\'e et al.  (1992) determined
that the radio-emitting region should be $\simgt 1$ AU in size.  Then we should
worry that, since we know that the source is variable in X-rays (see also below,
\S5.2), the radio-emitting plasma may live long enough against recombination
that it keeps a good chance to be detected in case the star-disk magnetic source
is cut off.  The recombination coefficient of a plasma having $kT_e > 0.05$ keV
is $\alpha_r = 3.5 \times 10^{-15} (Z+1)^2 I_Z^{1/2} (kT_e)^{-1}$ cm$^{-3}$
s$^{-1}$ ($Z$ = ion charge, $I_Z$ = ionization potential, energies in keV), so
that for hydrogen the recombination time is $t_r = 1/(\alpha_r n_e) = 6.1 \times
10^4$ s $\times (kT_e/$keV) $(n_e/10^{10}$ cm$^{-3})^{-1}$.  At the base of the
cone, near the star, the density is high and the recombination time is short;
however, for a constant velocity wind, $n_e(r) \propto r^{-2}$, and lower
densities dominate the emitting volume.  Taking $n_e = 5 \times 10^{10}$
cm$^{-3}$ at $r = R_{loop}$, we find at $\simgt 1$ AU $n_e = 2 \times 10^{7}$
cm$^{-3}$, and $t_r \simgt 3 \times 10^6$ s, or $\simgt 1$ month.

($iv$) Finally, travelling at $v_w \sim 300$\,km\,s$^{-1}$, the plasma will reach $r
\simgt 1$ AU in $\simgt 6$ days, i.e., more than 7 X-ray periods:  therefore
{\it we do not expect the radio emission of YLW15 to be variable on $\sim$ day
time scales, i.e., to carry any signature of the underlying stellar rotation}.
On the other hand, the emission will look continuous (or perhaps weakly
variable) as long as the magnetic shearing process giving rise to the X-rays
does not stop for more than $\simgt 1$ month at a time.  All we can say at this
point is that these conditions do not seem unreasonable.

We thus conclude that collimated hot coronal winds such as the one we have
invoked for YLW15 must contribute to the thermal radio emission of protostars,
and may perhaps even explain it.  In this case, what is commonly referred to as
``the ionized base of the outflow'' (or of the jets) to interpret the radio
emission, implying in particular temperatures $\sim 10^4$ K which are
unexplained so far, may rather be in reality AU-sized bipolar ``coronal winds''
at $\sim 10^6$ K, originating in magnetic interactions between the star and the
disk.

On the other hand, it may look surprising that we have not so far considered
{\it non-thermal} radio emission related to X-ray flares, since gyrosynchrotron
emission has been detected in a number of YSOs, and is directly tracing, like
the X-rays, their magnetic activity (FM).  However, it turns out that, contrary
to thermal emission, detections of non-thermal emission are extremely rare for
protostars.  The only {\it bona fide} case is that of the X-ray emitting Class~I
protostar R CrA IRS5, which has been seen with the VLA to emit a strongly
circularly polarized radio flux (Feigelson, Carkner, \& Wilking 1998).
Remarkably, the radio emission is variable on a time scale of $\sim 1$ day at a
level 0.15 -- 0.45 mJy.  This variability could perhaps be related to the
break-up period of the central object:  indeed, on the birthline the break-up
period is comprised between $\sim 13$ h and $\sim 33$ h for $M_\star \simlt 5
M_\odot$ (see \S4 and Fig.~6 below).  Gyrosynchrotron radio emission does have,
like X-ray flares, and contrary to the thermal emission, the potential to carry
the signature of some characteristic time scale tracing a stellar period.
However, this emission is easily absorbed by intervening fully ionized material
(including the coronal wind itself), and it seems that it can be detected only
in very favourable circumstances (temporarily lower mass loss, particularly
efficient electron acceleration, geometry, etc.).

	\subsection{Energy storage vs. energy release}
	
Turning to the {\it maximum} magnetic power which can be stored in 
the star-disk system with a magnetic structure explaining the observed 
flares of size $\approx r_D \approx r_A$, we have

\begin{equation} (dE_{mag}/dt)_{max} \sim \mid \Omega_\star- 
\Omega_D \mid \mu_\star^2/r_D^3 \approx 
\Omega_\star\mu_\star^2/r_D^3,
\end{equation}

\noindent (see, e.g., Shu et al.  1997).  With the above parameters for YLW15,
we find $(dE_{mag}/dt)_{max} \sim 10^{36}$\,erg \,s$^{-1}$.  Equivalently, over
one period of $\simlt 1$ day, $E_{mag} \simlt 10^{41}$\,erg can be released,
which shows that an enormous amount of magnetic energy is accumulated, at the
expense of the star-disk differential rotation.  The total energy released by
the ``superflare'' detected in YLW15 with {\sl ROSAT} by Grosso et al.  (1997)
is $E_{tot} \approx 10^{37-38}$\,erg, which suggests than such extremely
energetic events can {\it only} be explained by star-disk magnetic reconnection
phenomena.  There is of course no difficulty to explain lesser (but still
intense) flares like the quasi-periodic flares detected with {\sl ASCA}, for
which $E_{tot}$ is $\simlt 1$ order of magnitude smaller, since a large fraction
of the total luminosity may be emitted at lower temperatures in the form of
optical/UV/soft X-ray photons which would be more easily absorbed by intervening
material than at the $\sim$ few keV seen by {\sl ASCA}.  Also, another
fraction of the power will be kinetic:  for instance, the coronal wind discussed
in the preceding section has $dE_c/dt = \frac{1}{2} \dot{M}_{ej} v_w^2 \sim
10^{32}$ erg s$^{-1}$.  In addition, some of the power may go into particle
acceleration (see FM).

On the other hand, X-ray ``superflares'' of the intensity of YLW15 have never
been observed on any of the several hundred X-ray detected T Tauri stars, which
implies for them $(dE_{mag}/dt)_{max} = 0$ in Eq.~(14) above.  In turn, this
implies either $\Omega_\star = \Omega_D$, or $r_D \rightarrow \infty$.  The first
case must then apply to classical T Tauri stars (CTTS), while the second
corresponds to ``weak-line'' T Tauri stars (WTTS, or ``Class~III'' sources),
which have no disk (or at least no disk interacting with the central star).
Strictly speaking, this means that the only magnetic activity left for T Tauri
stars is their solar-like activity, which should then be essentially identical
for CTTS and WTTS, irrespective of the presence or absence of a disk.

This is fully consistent with observations:  the X-ray emission properties from
CTTS and WTTS are essentially indistinguishable, and their X-ray activity bears
all the symptoms of being solar-like, with no indication of the presence of a
disk playing any role (see FM).  More precisely, at least as far as X-ray
diagnostics are concerned, in spite of the fact that magnetic reconnections may
conceivably take place between the star and the disk, or even within the disk
itself (see FM; and also Aly \& Kuijpers 1990), observations of classical T
Tauri stars show that these reconnection events must be much less important than
stellar solar-like magnetic activity.  (However, it cannot be excluded that such
events generate lower-temperature flares, which would emit EUV radiation easily
absorbed by the intervening material, and/or be undetectable by X-ray
satellites.)

We conclude that extreme X-ray luminosities are never released at the T Tauri
stage because a significant change of star-disk magnetic configuration leading
to reconnection has become impossible:  in the case of CTTS, the magnetosphere
and the inner boundary of the disk are somehow bound together in solid rotation
as a result of magnetic braking, and possible magnetic interactions within the
disk beyond this boundary must be comparatively small.  At this later stage,
magnetic reconnection events leading to the observed X-ray flares are therefore
only those resulting from convective movements at the stellar surface, as they
are in the case of WTTS, where the disk is absent.

\section{WL6: a slowly rotating protostar}

Kamata et al.  (1997) present an {\sl ASCA} observation of another Class~I
protostar in the $\rho$ Oph cloud, WL6 (Wilking \& Lada 1983), also known as
YLW14, spread over $\sim 20$ h.  They give a sinusoidal fit to the light curve,
but actually only about 3/4 of such a sinusoid is seen.  These authors propose
that the variation is really periodic, with a period $P_\star = 0.97$\,day, and
attribute it to a rotating large active region at the stellar surface.  We note
however that a strictly sinusoidal behavior is not compatible with an eclipse
phenomenon, whether it refers to a rotating active region, to a binary system,
or more generally to some external occultation (see for instance the periodic,
but highly non-sinusoidal X-ray light curve of the O7 star $\theta^1$C Ori
interpreted in terms of disk occultation by Babel \& Montmerle 1997).

We have therefore reexamined the observation of Kamata et al.  in the light of
the recent work on rotational modulation of X-ray flares by Stelzer et al.
(1999), which gives a new interpretation of ``slowly'' rising and decaying light
curves similar to that observed in WL6.  The idea, first proposed by Casanova
(1994) for the $\rho$ Oph CTTS SR12, is that an X-ray flare has occurred on the
back side of the star prior to the beginning of the observation, and that the
observation starts during the cooling phase, while the magnetic loop confining
the plasma appears above the limb and comes into view.  For WL6, the X-ray
temperature behavior along the light curve apparently does not change
significantly, being successively (after rebinning into two time intervals)
$kT_X$(keV) $=2.2(+0.5/-0.4)$, and $2.2(+1.0/-0.6)$, but the uncertainties are
large, so we concentrate on the light curve.  We model it as in Stelzer et al.,
with 6 parameters:  maximum intensity, quiescent level, e-folding cooling time,
start of flare egress above the horizon, stellar period, and radius of the loop.
Figure~\ref{WL6} shows the result.  We find that the data can be fit in
particular with a comparatively small equatorial plasma bubble (radius/$R_\star
\sim 0.8$), and a {\it slow rotation} ($P_\star = 3.3$\,days), much slower than
inferred from the sinusoidal fit.  The reduced $\chi^2$ is 1.3 for 12 degrees of
freedom, which is comparable to the sinusoidal fit ($\chi^2= 1.1$ for 14 degrees
of freedom).  The light curve of WL6 can be compared to the light curve of YLW15
(see Fig.~\ref{YLW15}).

\vspace{0.25cm}
\centerline{\hbox{\psfig{file=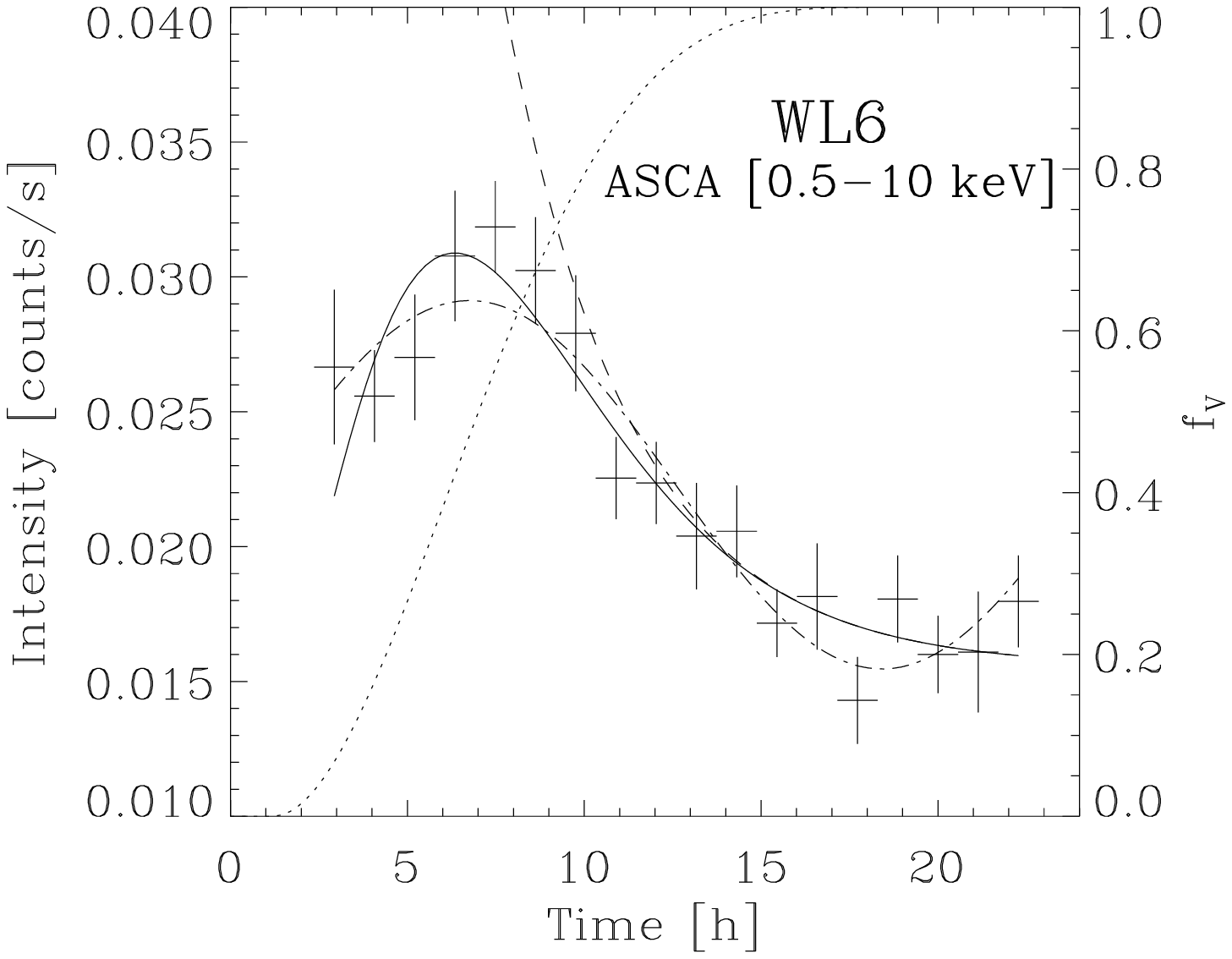,width=0.45\textwidth}}}
\vspace{-0.25cm}
\figcaption{\small Light curve of WL6 obtained with {\sl ASCA} (Kamata et al.  1997),
reexamined in the framework of the rotational modulation of a spherical flare
starting on the back of the star (Stelzer et al.  1999).  The fitted parameters
of the continuous line are:  quiescent level (= 0.016 cts s$^{-1)}$; flare
maximum (= 0.110 cts s$^{-1}$); flare cooling timescale ($\tau = 3.6$\,h);
moment of egress after the start of the observation ($t_{egress} = 1.2$\,h);
plasma volume fraction $f_V$ (= visible volume/final volume) as a function of
time, as seen by the observer, indicated by the dotted line; stellar period
($P_\star = 3.3$\,days); and ratio between the plasma sphere radius and the
stellar radius ($f = 0.76$).  The reduced $\chi^2 = 1.3$ for 12 d.o.f.  The
dashed line shows the flare cooling phase without the rotational modulation.
The dot-dashed line is the sinusoidal fit by Kamata et al.  1997 (reduced
$\chi^2 = 1.1$ for 14 d.o.f.).  Note the difference with YLW15 (see
Fig.~\ref{YLW15}) in terms of luminosity (count rate on the left vertical scale)
and duration (horizontal scale).\label{WL6}}
\vspace{0.25cm}

At this point, it is important to emphasize that while being also classified as
a Class~I protostar on the basis of its IR properties, WL6 is a quite different
object from YLW15.  Situated near the B1/B2 core region, it is not embedded in a
large condensation like YLW15, but rather appears isolated (MAN).  Its
bolometric luminosity is $L_{bol} \sim 1.1$\,L$_\odot$ (Lada \& Wilking 1984).
While its rising slope at short wavelengths, with $\alpha_{IR}(2-10 \mu$m) =
1.0, formally makes it a Class~I protostar, its SED is essentially flat beyond
3.4\,$\mu$m, so that it rather belongs to the ``flat-spectrum'', or embedded T
Tauri, category.  In fact, a classical T Tauri star with an edge-on disk could
perhaps give a better interpretation of this kind of SED (see, e.g., Chiang \&
Goldreich 1999), and would give additional support to our interpretation of the
X-ray light curve in terms of an equatorial rotationally modulated flare.  No
circumstellar envelope could be detected in the millimeter observations by AM
and MAN:  this yields an upper limit to the envelope mass $M_{env} \simlt
10^{-3}$\,M$_{\odot}$.  An outflow weaker than the one of YLW15 has been
detected by Sekimoto et al.  (1997).  From these characteristics, {\it WL6
appears as significantly more evolved than YLW15}.  In addition, no VLA
emission has been detected, down to a very low limit ($F_{5GHz} < 100~\mu$Jy,
i.e., $\sim 30$ times less than for YLW15).  In spite of its isolation away
from the $\rho$ Oph dense cores and its small envelope, its extinction,
classically derived from near-IR data, is high:  taking the recent near-IR data
by Barsony et al.  (1997) and assuming a flat-spectrum disk with an intrinsic
$H-K = 0.9$, we find $A_{V,IR} \sim 40$; a lower value has been found from the
X-rays:  $A_{V,X} \sim 20$.  This would be consistent with an edge-on disk,
again invoking a lower gas-to-dust ratio than usual (grain condensation in the
disk ?).  Uncertainties on $A_V$ are clearly large anyway.

Choosing $A_{V} \sim 20$ for consistency with the X-ray data, the maximum
(hidden) luminosity given by the rotationally modulated flare model is then
$\sim$ 3.8 times higher than the value derived by Kamata et al., to become
$L_{X,max}$[2-10\,keV] $\sim 5.3 \times 10^{31}$\,erg\,s$^{-1}$.

This luminosity value is high compared to the average X-ray luminosity of T
Tauri stars ($L_X \simeq 3 \times 10^{28} - 3 \times 10^{29}$\,erg\,s$^{-1}$,
see FM), and only 4--20 times less luminous than the unusually strong X-ray
flare recently observed on the weak T Tauri star V773 Tau (Skinner et al.  1997;
also studied by Stelzer et al.  in terms of rotational modulation), for which a
comparable rotation period is obtained.  This suggests that the central star of
WL6 behaves in X-rays more like an ``extreme'' T Tauri star, rather than as a
protostar like YLW15, owing to its much slower rotation.  {\it This is
consistent with its status independently deduced from IR and mm observations.}
In that case, the WL6 X-ray flare would be due to (enhanced) solar-like activity
on its surface, rather than to star-disk interactions.  This conclusion is
also consistent with the absence of detectable cm radio emission.

\section{Stellar parameters for YLW15 and WL6}

In the preceding sections, we had to use estimates of stellar parameters such as
the radius or the mass, and the accretion rate.  {\it Stricto sensu}, at the
Class~I stage these quantities are rather ill-defined theoretically and cannot
be measured directly.  Class~I characteristics are derived from the envelope
(mainly via its SED), not by the central object.  However, the X-ray behaviour
of YLW15 and other X-ray detected protostars strongly suggest that a genuine
stellar object, with a photospheric radius and a mass, is present inside the
circumstellar envelope of Class~I protostars.

To refine our previous estimates, we therefore make the step to consider that
YLW15 and WL6 are very close to their ``birthlines'' (e.g., Stahler 1988) where
they become optically visible, at least as far as the {\it stellar} properties
are concerned.  Even if the envelope properties of YLW15 suggest that it has not
quite reached that stage, the envelope mass determined from mm measurements
(\S1) is $\ll 1 $\,M$_\odot$, which implies that the central object has indeed
essentially reached its final mass.  This is even more true for WL6, which does
not have a detectable envelope.  

We can then use the models by Palla \& Stahler
(1992, 1993; hereafter PS) which define a ``protostar mass-radius relation''
(hereafter the PS relation) as a function of the accretion rate
$\dot{M}_{acc,\star} \sim a^3/G$ (where $a$ is the sound speed in the dense
core; e.g., Shu, Adams, \& Lizano 1987) building up the star.  The standard
value (for dense cores with temperatures $\sim 30$\,K) is $\dot{M}_{acc,\star}
\sim 10^{-5}$\,M$_\odot$ yr$^{-1}$ (Adams, Lada, \& Shu 1987).  

\vspace{0.25cm}
\centerline{\hbox{\psfig{file=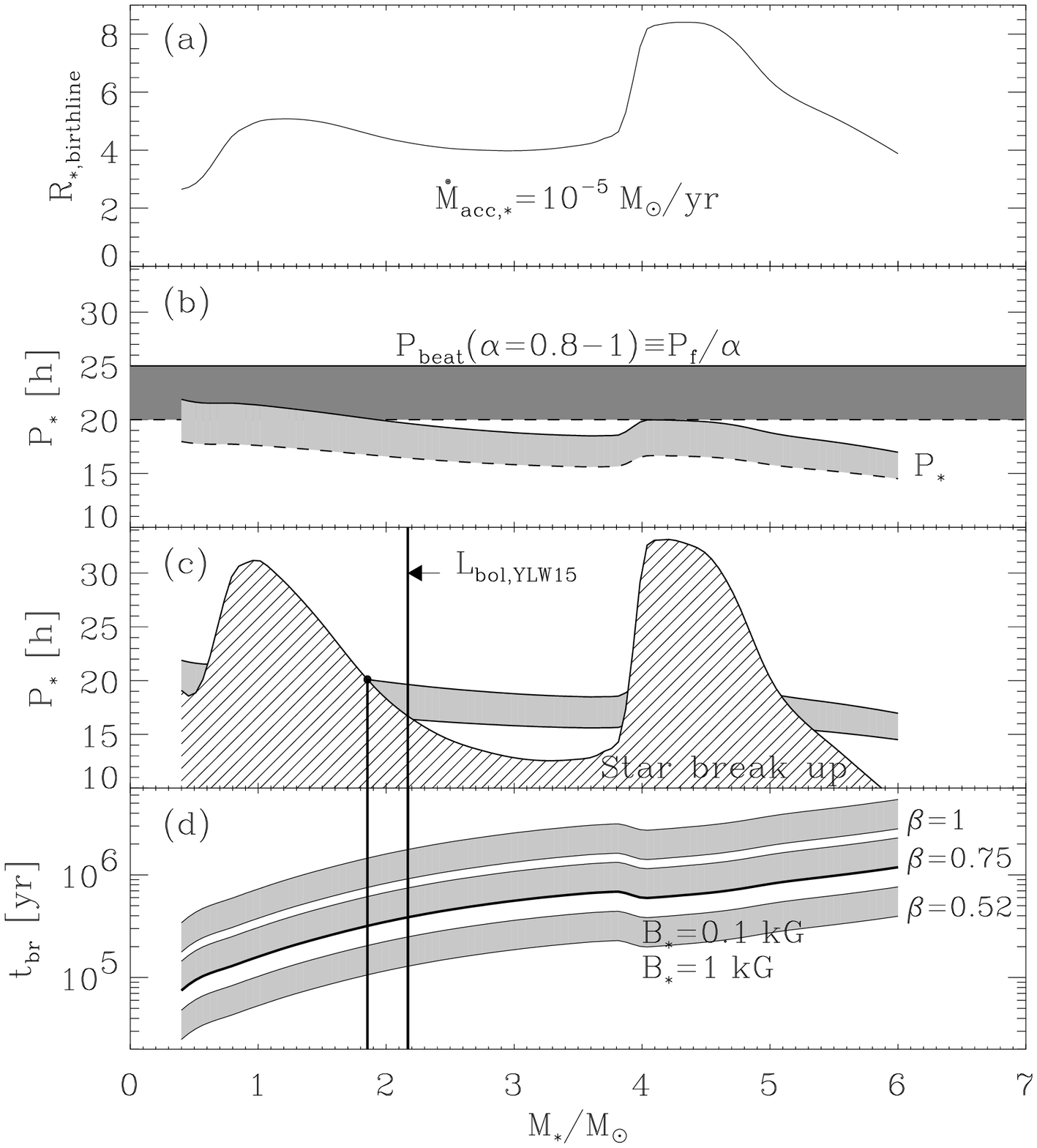,width=0.51\textwidth}}}
\vspace{-0.25cm}
\figcaption{\small ($a$) Mass-radius relation assuming a star build-up accretion rate, $\dot{M}_{acc,\star} = 10 ^{-5}$\,M$_\odot$/yr, on the ``birthline'' (Stahler 1988 for $M_\star < 1.5$\,M$_\odot$; Palla \& Stahler 1992 for higher masses). ($b$) Relation between the rotation period of YLW15, $P_{\star}$ (calculated from the flare period, the magnetic loop size of the YLW15 X-ray flares, and the mass-radius relation shown on the first panel), and the beat period, $P_{beat}$, deduced from the flare period, assuming the relation $P_{beat}=P_f/\alpha$ (dashed line: $\alpha=1$; continuous line: $\alpha = 0.8$). The grey strip shows the computed values of the star rotation period for $\alpha = 0.8 - 1$ (see text, \S 2.1.3; same strip as shown in Fig.~4). ($c$) Corresponding break-up period as a function of stellar mass, and forbidden periods underlined by the shaded area (period shorter than the break-up period). The right vertical line corresponds to the upper limit to the mass of YLW15 deduced from its observed bolometric luminosity (see text, \S 4); the left vertical line is the lower limit derived from the rotation constraints (panel 3). ($d$) Braking time $t_{br}$ as a function of mass (and radius via the mass-radius relation shown on panel $a$) for a disk accretion rate $\dot{M}_{acc} = 10^{-6}$\,M$_\odot$/yr, three values of $\beta$ describing the accretion geometry ($\beta = 0.52$ for a geometrically thin accretion disk, $\beta = 1$ for spherical accretion), and two values of the photospheric magnetic field $B_\star$ (0.1\,kG, and 1\,kG). For reference, the duration of the Class~I stage is estimated at $0.75-1.5 \times 10^5$\,yr (Luhman \& Rieke  1999). The thick curve  ($B_\star \sim 1$ kG, $\beta \sim 0.75$) is our ``best choice'' to explain at the same time the fast rotation of YLW15 and the slow rotation of WL6. We see that the braking time is then shorter than the age of Class~I protostars for masses smaller than $\approx 1-2 M_{\odot}$. \label{period}}
\medskip

PS also find independently that a good fit to the observed locations of the youngest low- to
intermediate- mass stars in the H-R diagram is obtained with this value of
$\dot{M}_{acc,\star}$.  PS then derive two series of models, depending on the
accretion mode, spherical or from a disk:  given the young stage of YLW15 and
WL6, we take the disk-accretion build-up model, and we will therefore use the
corresponding PS relation (shown in Fig.~\ref{period}$a$) in what follows.

On the other hand, outflow observations show evidence for a strong decrease in
the accretion rate from the build-up stage (i.e., Class 0) to the Class~I
stage (end of build-up) (Bontemps et al.  1996), as shown by their
evolutionary diagram plotting $(F_{CO}/L_{bol}) vs.  (M_{env}/L_{bol})^{0.6}$,
where $F_{CO}$ is the outflow momentum flux.  Such observations are supported by
some outflow models (see Henriksen, Andr\'e, \& Bontemps 1997 and references
therein).  In this diagram, the positions of YLW15 and WL6 suggest
$\dot{M}_{acc} \approx 10^{-6}$\,M$_\odot$/yr.  
A low accretion rate of this order is also required at the Class~I stage 
in order that the accretion luminosity $L_{acc} = G M_\star \dot{M}_{acc}/R_\star$ 
does not exceed the observed bolometric luminosity ($L_{bol} \sim 10$\,L$_\odot$ for YLW15).

Let us discuss YLW15 in more detail.  
($i$) Having the X-ray flare period, the size of the magnetic loop and the 
PS mass-radius relation, we find from
Eq.~(\ref{eq:periode_etoile}) that the star rotation period 
must be smaller than the beat period (\S 2.1.3 above; see Fig.~\ref{period}$b$).  
($ii$) The equatorial velocity of YLW15 cannot exceed the break-up velocity,
which translates into a limiting condition on the star rotation period,
i.e., one must have $P_\star > P_{\star,0}$ (break-up period, 
illustrated in Fig.~\ref{period}$c$), with permitted zones
in the mass ranges $M_\star \sim 1.8 - 4.0$\,M$_\odot$ and 
$M_\star \simgt 5$\,M$_\odot$, as indicated by the hatched area.   
($iii$) An independent constraint comes from the observed
$L_{bol}$ taken at the birthline:  the photospheric luminosity
$L_\star = 4\pi R_\star^2 \sigma T_{eff}^4$ should be at most equal to $L_{bol}$
(any difference being attributable to the accretion luminosity):  
from the birthline on the H-R diagram (Palla \& Stahler 1993), 
we find $M_\star \leq 2.2$\,M$_\odot$ 
(see Fig.~\ref{period}$c$).
Combining all these constraints, we conclude that the mass and radius
of the central object of YLW15 must lie in the ranges
$M_\star=1.8 - 2.2$\,M$_\odot$, and $R_\star = 4.6-4.2$\,R$_\odot$ 
(PS relation).  
In other words, according to the above arguments, YLW15 must be {\it a future
Herbig AeBe star}, itself the progenitor of an A star. It is noteworthy that
this status then sets YLW15 apart from the typical, lower-mass YSO in the $\rho$ Oph cloud,
a fact that was already noticed in relation with some of its particular
properties described in \S1.

On the other hand, WL6, by way of its small luminosity ($L_{bol} =
1.1$\,L$_\odot$) is necessarily a low-mass star ($M_\star \simlt
0.4$\,M$_\odot$), with $R_\star \simlt 2.7$\,R$_\odot$ (PS relation).  Returning
to the braking times (Eq.~(\ref{rayon_alfven})), and adopting the same values of
$B_\star$, $\dot{M}_{acc}$ and $\beta$ as for YLW15, we find that for WL6
(assumed initially rotating at break-up, i.e., $P_{\star,0} \sim 19$\,h from PS)
$t_{br,0}$ is at least 5 times ($\sim$ the mass ratio) shorter than for YLW15.
As a result, WL6 must have gone through most of its braking phase, so that its
magnetic field is by now likely locked to the accretion disk.  Comparing the
rotation characteristics of YLW15 and WL6, we conclude that {\it YLW15 is a fast
rotating protostar of intermediate mass, while WL6 is a spun-down low-mass
protostar.}

\section{Discussion}

\subsection{A mass-rotation relation close to the birthline ?}

More generally, Fig.~\ref{period}$d$ displays the magnetic braking timescale
from initial break-up (Eq.~9) as a function of stellar mass.  If the accretion
rates and magnetic fields are the same for the high and the low stellar masses,
we find that the rotational status of protostars is mainly governed by the
stellar mass, unless other factors like $B_\star$ or the accretion geometry
parameter $\beta$ are very different from star to star.  The X-ray observations
certainly suggest that $B_\star$ should not vary much.  Fig.~6$d$ illustrates
all the possible cases.  However, using X-ray flaring as a clock, we have
``measured'' the rotation periods of YLW15 (fast) and WL6 (slow), and we have an
estimate of their masses.  In the absence of a better individual age
determination for these protostars than an average value $\sim 1-2 \times 10^5$
yrs, we can then use their rotation properties to ``normalize'' the set of
$t_{br} - M_\star$ curves of Fig.~6$d$:  we find that $B_\star \sim 1$ kG and
$\beta \sim 0.75$ is a reasonable combination (thick line).  The value $B_\star
\sim 1$ kG is fully compatible with all the X-ray observations of YSOs (FM); the fact
that $\beta \sim 0.75$, could be an indication that for Class~I protostars the
actual accretion geometry is a mixture of disk ($\beta = 0.52$) and spherical
accretion ($\beta = 1$).

In these conditions, the critical mass $M_{crit}$ such that, near the birthline,
stars with $M_\star > M_{crit}$ may be fast rotators (i.e.  $t_{br,0} >$
lifetime) is $M_{crit} \approx 1 - 2 M_\odot$.  This limit may not be very
precise, however, because low-mass stars are clearly more affected by
star-to-star variations of the various parameters entering Eq.~(9), and also
because individual stellar lifetimes are still uncertain.

This conclusion may be affected at a later stage during the evolution by a
change in the accretion rate, suggested by outflow observations (see \S 4).  For
low-mass objects like WL6, which are braked quickly and are now slowly rotating,
a late decline in the disk accretion $\dot{M}_{acc}$ ($\sim 10^{-7} M_\odot$
yr$^{-1}$ at the T Tauri stage) has no consequence.  However, like YLW15
higher-mass objects may end their Class~I stage while still appreciably
rotating:  since $t_{br}$ is approximately inversely proportional to
$\dot{M}_{acc}$, the braking effect will essentially stop just before the
birhline, and the star will continue to spin down but on much longer timescales.
In other words, even some time after the birthline the rotation status will
reflect the conditions on the birthline:  taking into account the fact that the
stellar mass turns out to be the most widely variable parameter in Eq.~(9), our
estimate of the magnetic braking time scale predicts a {\it mass-rotation
relation} of the form $t_{br} \simpropto M_\star$, where close to the
birthline the stars should be slow rotators below $M_{crit} \approx 1 - 2
M_\odot$, and fast rotators above. Such a relation can also be considered as
an initial condition for the rotation history of T Tauri stars.

Observationally, this could be an explanation for the existence of rapid
rotators among very young stars:  our prediction here is that rapid rotators
should in general be relatively massive.  There are some hints that this may be
the case:  the three rapid rotators ($P_\star = 1.5-2$\,days) among the list of
classical T Tauri stars with measured rotation periods by Bouvier (1990) all
have masses between 1.5 and 2.0\,M$_\odot$, and Herbig AeBe stars, which have
masses $\sim 2 - 8 M_\odot$, seem, as a rule, to be fast rotators (Boehm \&
Catala 1995).

\subsection{Fast stellar rotation vs. accretion}

An intriguing paradox arises if the central star of YLW15 is really rotating
near break-up.  Indeed, in such conditions the presence of a strong magnetic
field (enough to balance the incoming accretion flow at $r_A$) would inhibit
accretion onto the central object and convert all the incoming material into an
outflow, thereby preventing the star to form in the first place.  This shows
that accretion and magnetic interactions cannot be steady phenomena:  episodes
dominated by star-disk magnetic interactions and (periodic) flaring must be
separated by episodes of intense accretion, accompanied by different ejection
phases.  The succession of intense accretion episodes is reminiscent of the FU
Orionis phenomenon, in which the accretion rate is so high ($\dot{M}_{acc}
\approx 10^{-4} M_\odot$ yr$^{-1}$) that the star is engulfed by the disk
(e.g., review by Hartmann, Kenyon, \& Hartigan 1993).  The corresponding duty
cycle for protostars is unknown, but in the case of YLW15, we have already seen
two different flaring regimes:  a ``high state'', with a superflare and a
quasi-periodic flaring episode, as well as a ``low state'', with a non-detection
by {\sl ASCA} on one previous occasion (Koyama et al.  1994; Kamata et al.
1997).

On the other hand, the existence of ``high'' and ``low'' X-ray activity states
is reminiscent of the QPO phenomenon, and indeed the suggestion that episodic
flaring may occur as a result of reconnection between a neutron star and its
accretion disk has been suggested in the past (see Aly \& Kuijpers 1990).
Alternatively, such a phenomenon might be explained in the framework of
accretion instabilities in magnetized disks (Tagger \& Pellat 1999).  Other
mechanisms may be invoked.  For instance, it is known that X-rays ionize the
accretion disk and, at least in T Tauri stars (where $\dot{M}_{acc} \approx
10^{-7}$\,M$_\odot$\,yr$^{-1}$), may regulate the accretion rate via the
(small-scale) Balbus-Hawley magnetorotational instability (Glassgold, Najita, \&
Igea 1997; Glassgold, Feigelson, \& Montmerle 2000). One can then imagine a
feedback mechanism based on the X-ray flares themselves:  
more X-rays means more ionization, and thus more accretion, which, if high
enough, may engulf the magnetic field and stop the flaring, or at least absorb
the X-rays near the star, so that the disk ionization drops, slowing the
accretion flow and letting the X-ray flaring phase resume to start a new cycle.

Obviously this situation is new and complex, and detailed 3D calculations are
needed to take into account the accretion, the magnetic field deformation
preceding the reconnection, and possibly the feedback of X-rays on accretion.
The {\sl ASCA} observation of a triple X-ray flare reported in Paper I clearly
forces us to reconsider the details of the mechanisms of accretion in forming,
fast rotating stars.

\subsection{Class~I protostars: X-ray loud vs. X-ray quiet}

The preceding discussion also suggests that rather stringent conditions must be
met for a Class~I protostar to have detectable X-ray emission.  Given the high
extinctions, the detectable X-ray luminosities are $L_X \simgt 10^{30-31}$\,erg
\,s$^{-1}$ at the level of sensitivity of {\it ROSAT} or {\it ASCA}, which is
higher than the values corresponding to the usual solar-like activity seen on T
Tauri stars.  Thus, as argued in \S2.4, it is likely that very high X-ray
luminosities can be produced {\it only} as a result of star-disk magnetic
reconnection phenomena.  The example of YLW15, the only one so far for which we
have evidence for superflares and periodic flaring, suggests that the underlying
object must be a fast rotating star (i.e, near break-up) not yet
magnetically locked to its accretion disk.  In contrast, if this object is a
slowly rotating star (i.e., far from break-up), its solar-like surface
activity may be detected only in favourable circumstances, as in the case for
WL6.

More generally, in addition to YLW15 and WL6, about a dozen protostars have been
detected in X-rays so far, by {\sl ROSAT} and/or by {\sl ASCA}, with
luminosities $L_X \simgt 10^{30-31}$\,erg\,s$^{-1}$ (FM).  
Many, but not all, display X-ray flares, with
time scales quite comparable to those observed on T Tauri stars, possibly
hotter ($T_X \sim$ a few keV) and more luminous.  All are Class~I and
flat-spectrum protostars.  However, the number of ``X-ray loud'' Class~I
protostars (including in $\rho$ Oph El29, IRS44 and IRS46) is only a few percent 
of their total number (Carkner et al.  1998).
Sekimoto et al.  (1997) have proposed that an orientation effect could be the
reason, arguing that X-ray emitting protostars must the ones seen pole-on
because of the lower extinction along the funnel created in the envelope by
molecular outflows.  However, in the case of YLW15, the {\sl HST} image
clearly indicates that the accretion disk is viewed almost edge-on, and
at least for this source an alternative interpretation is necessary.

In the context of the present paper, we can offer other reasons why the majority
of Class~I sources have not been detected yet in X-rays.  First, reconnection
phenomena may not necessarily occur on a regular basis, for instance because of
various possible accretion instabilities, as discussed in the preceding
subsection.  (Recall that YLW15 itself was not detected during one previous {\sl
ASCA} observation.)  Second, if its mass is low enough, the star may have
already braked early in the Class~I stage ($t_{br} < 10^5$\,yrs) to reach
corotation with the inner accretion disk.  The very youngest may be X-ray
bright, but in most cases only an underlying solar-like activity will go on, at
levels lower than for WL6 for instance.  

An intringuing property can also be explained in this context.  To date, {\it
all} the X-ray detected protostars belong to clusters (in addition to $\rho$
Oph:  RCrA, Chamaeleon, Orion), whereas none has been detected in Taurus, for
example.  We suggest that this is because clusters are the only place where
stars can be found to be sufficiently massive ($M_\star > 1$\,M$_\odot$) to be
fast rotators.  Only lower-mass stars form in diffuse regions like Taurus,
where the average stellar mass is more like $M_\star \sim 0.5$\,M$_\odot$:  most
protostars in such regions would presumably have reached magnetic locking by
now, and therefore be the seat of an as yet undetectable low-level solar-like
X-ray activity.

	\subsection{Areas of uncertainty}
	
The preceding sections have shown that it is possible to construct a
self-consistent model to explain the periodic X-ray flaring observed in the
Class~I protostar YLW15 by {\sl ASCA} in terms of a reconnection within a
magnetic loop sheared between a fast rotating star and a slowly rotating
accretion disk.  Thanks to this approach, it has been possible for the first
time to recognize the fast-rotating status of a protostar, and propose an
explanation for the existence of extremely luminous X-ray flares and other X-ray
properties of protostars.

Now it must be realized that this construction rests on foundations that
are not all certain, and more work is needed to confort them.

\begin{itemize}

\item{The exact topology of magnetic field loops confining the X-ray emitting
plasma during the decay phase is often subject to debate, since X-ray
observations give access to the emission measure only ($EM = \int{}{} n_e^2
dV$), and examples are known (including on the Sun) of reheating phases during
cooling.  In the case of the triple flare of YLW15, however, the models with
radiative cooling do not call for departures from a simple (if admittedly
approximate) semi-circular magnetic loop.  Thus, we consider the loop radius
$R_{loop}$, which in our model plays an essential role to determine the beat
period between the star and the disk, and thus {\it in fine} the stellar period,
as reasonably well established.  Data of better quality, such as provided by the
uninterrupted observations of {\sl XMM} or {\sl Chandra} will put more
constraints on such models in the future.}

\item{Although recent theoretical progress on MHD simulations of sheared
magnetic configurations has been quite substantial, much remains to be done,
especially in 3D, to even start comparing the results with ``real''
magnetic configurations resulting from star-disk shearing.  We have used the
numerical results obtained by Amari et al.  (1996) for the time evolution of a
twisted magnetic tube to infer that reconnection in the case of star-disk
magnetic shearing should not occur much before one beat period, but of course
actual calculations are required to validate this inference.  }

\item{The ``classical'' calculation of magnetic braking rests on the model
initially proposed by Ghosh \& Lamb (1978).  This model assumes a steady-state
solution to the problem of finding the balance between the star-disk magnetic
torque and angular momentum variation.  But as discussed above, at least in the
case of fast rotation a steady-state situation likely does not hold; a possible
manifestation is, precisely, magnetic shearing followed by reconnection and
flaring on stellar rotation time scales.  In addition, the interaction between
the stellar magnetosphere and the inner disk remains poorly known, and the
momentum loss mechanism is unclear.  Our calculations of the corresponding
braking time $t_{br}$ are meant to represent the long-term evolution, and should
still be taken as a rather rough approximation.  Whether a collimated hot
coronal wind actually contributes to the stellar angular momentum loss is also a
possibility, but our calculations of the mass-loss rate (and of the inferred
thermal radio emission) should be considered merely as a starting point.
Nevertheless, the bottomline is that $t_{br}$ seems to be of the same order as
the duration of the Class~I protostar phase, so that some of these protostars
may still be rotating fast now, while others may have substantially spun down.}

\item{The ``mass-rotation'' relation we have advocated, and in particular the
limit between fast and slow rotators near the birthline, is correspondingly
uncertain.  Our calculations predict this limit should lie around $\approx 1-2
M_\odot$, and existing observations tend to support this.  However, too little
is known observationnally on the rotation of stars in this mass range and higher
to be more conclusive:  additional data must be obtained near the birthline,
especially among the Herbig AeBe stars.  The exact accretion geometry also plays
a role and this is a concern as well.}

\end{itemize}

Given these uncertainties, the present work may be considered as
encouraging, but still exploratory.

	\section{Summary and Conclusions: \\
A tentative history of protostellar rotation and magnetic activity}

We have studied the implications of an intense hard X-ray ``triple flare''
observed with {\sl ASCA} at the level of $L_X \sim 10^{32}$ erg s$^{-1}$ on the
Class I protostar YLW15 in the $\rho$ Oph cloud (analyzed in Paper I).  With the
reasonable assumption that magnetic shearing followed by reconnection between
the central star and the accretion disk is the basic heating mechanism for these
flares, we find that the observed flare quasi-periodicity of $\sim$ 20\,h,
probably also observed over 10 years ago by {\sl Tenma}, is tracing the fast
rotation of the central star. A key parameter in the demonstration is the size
of the X-ray emitting region, which we model as a semi-circular magnetic loop
of radius $\sim 4.5 R_\odot$.

Given the evolved stage of Class I protostars, we further assume that the
central star is on the ``birthline'' of the H-R diagram, so that a
relation exists between its mass and its radius.  Various constraints led us to
the conclusion that the central star of YLW15 has $M_\star = 1.8 -
2.2$\,M$_\odot$ and $R_\star = 4.6-4.2$\,R$_\odot$, i.e., is a future A star.
Another $\rho$ Oph Class I protostar, WL6, more evolved than YLW15, and also
detected by {\sl ASCA}, displays an X-ray light curve which we interpret
in terms of slow rotation, with a period $\sim 3$ days; its low bolometric
luminosity places an upper limit to its mass of $M_\star \sim 0.4$\,M$_\odot$,
and thus $R_\star \simlt 2.7$\,R$_\odot$.

On the other hand, we have found that magnetic braking likely takes place on
timescales of a few $10^5$ yrs, i.e., {\it within the Class~I stage}, so that
both fast and slowly rotating protostars may coexist in the same star-forming
region.  These results obtained for YLW15 (fast rotation) and WL6 (slow
rotation) support this finding, but are obviously subject to future confirmation
(in particular from observations of other protostars with the upcoming
generation of X-ray satellites).  YLW15 and WL6 however belong the same
star-forming region and are well documented at many wavelengths, which adds
weight to the conclusions.  We also found that {\it the main parameter for
braking is the stellar mass}, since X-rays always indicate comparable values of
the magnetic field ($\approx 1$\,kG at the photospheric level).  The accretion
rate is also important, but a value of $\approx 10^{-6}$\,M$_\odot$\,yr$^{-1}$
during the Class~I protostar stage seems to be typical and fairly independent of
the mass (for low- to intermediate-mass stars).  With a duration of $\simlt
2\times 10^5$\,yr for the Class~I stage, we find that the boundary between fast
and slow rotators at the birthline is $\approx 1 - 2$\,M$_\odot$.

More generally, since all the available X-ray evidence indicates that 
stellar rotation and magnetic activity are intimately 
related in low-mass stars, we may now put these results in a broader 
context. Moving backwards in time, we may propose the following scenario for 
the rotational evolution of protostars:

(i) At ages $\sim 10^5$ yrs, Class~I protostars can be envisioned as Class~II T
Tauri stars embedded in a circumstellar envelope containing more mass when the
star is younger.  Their magnetic field, which may be primordial or
dynamo-generated, exerts a braking torque on the accreting material in the
vicinity of the central, growing star.  Since the braking time is typically a
$\le$ few $10^5$ yrs, T Tauri-like synchronous rotation with the disk may be
reached {\it during} the Class~I stage, the exact moment depending on parameters
like the mass (mainly) and the surface magnetic field and accretion rate and
geometry (to a lesser extent).  Young, fast rotating (and hence presumably more
massive) Class~I sources will be strong X-ray emitters due to star-disk
interactions (YLW15), while evolved, less massive slow rotators will exhibit
only solar-like X-ray activity near the surface of the central star (WL6).
Fast-rotating Class~I protostars may also be thermal radio emitters (from their
hot coronal winds), and under very favourable circumstances, non-thermal radio
emitters (from flaring activity).

(ii) At the early, build-up protostar stage (Class 0, ages $\simlt 10^4$ yrs;
see AWB), there is currently no observational indication of rotation:  indeed,
the fact that no Class~0 protostar has been detected in X-rays so far can easily
be attributed to their huge extinction ($A_V$ up to $\sim 1000$) due to their
massive envelope.  Radio emission does not suffer from this drawback:  the
prototype Class~0 source, VLA1623, was first identified as a radio source.
Thermal radio emission, one of the signatures of the Class~0 stage, may
indirectly trace magnetic activity via a coronal wind, but as discussed in
\S2.4, it will probably not carry any rotation signature, contrary to the X-rays
and possibly the non-thermal radio emission.  Based on the Class~I results
presented in this paper, we can only speculate that the central object initially
acquires angular momentum from its envelope and/or accretion disk, and builds up
during the Class~0 phase in a state of fast, near break-up rotation.  As is now
well documented, outflows appear as soon as protostars exist (Andr\'e,
Ward-Thompson, \& Barsony 1993; Bontemps et al.  1996).  This can be due among
others to an infall-ejection mechanism, guided by magnetic field lines
(Henriksen \& Valls-Gabaud 1994; Fiege \& Henriksen 1996), or by a
magnetocentrifugal mechanism reminiscent of the first version of the X-wind
model proposed by Shu et al.  (1988), which was based on break-up rotation of
the central star.  In such models, the magnetic field of the central object may
be simply of interstellar origin and dragged during collapse.  Reconnections
followed by coronal mass ejections or a coronal wind may also play a role in
launching YSO jets and/or outflows.  Whatever the mechanism, it probably cannot
be in a steady state if magnetic fields are important, in order to reconcile
fast rotation and accretion.

The detection of X-rays from Class 0 protostars would of course be crucial in
revealing their magnetic fields and estimating their strength, in tracing the
rotation of the central, growing star, and in understanding the relations
between accretion and outflow mechanisms.  This will soon become possible up to
$A_V \simgt 100$ with {\sl XMM} owing to its high effective detection area.  To
clarify the relative importance of coronal and disk winds will require the help
of sensitive very high-angular resolution radio interferometers.  This class of
observatories offer unprecedented opportunities to probe magnetic fields as an
essential ingredient of star formation.

\acknowledgements 
We thank Jean-Jacques Aly, Tahar Amari, Philippe Andr\'e, Eric
Feigelson, Kensuke Imanishi, Francesco Palla, Michel Tagger, and Yutaka Uchida
for many useful discussions in the course of this work. We also thank an
anonymous referee for stimulating remarks which helped to improve the clarity
of the paper.

\end{document}